\documentclass[preprint,12pt]{elsarticle}

\usepackage{hyperref}
\usepackage{caption}
\usepackage{tabularx}
\usepackage{array}
\usepackage{multirow}
\usepackage{graphicx}
\usepackage{caption}
\usepackage{subcaption}
\usepackage{multirow}
\usepackage{multicol}
\usepackage{geometry}
\usepackage{longtable}
\usepackage{booktabs}
\usepackage{url}
\usepackage{lscape}
\usepackage{xcolor}
\usepackage{comment}
\usepackage[T1]{fontenc}

\newcounter{subsubsubsection}[subsubsection]
\renewcommand\thesubsubsubsection{\thesubsubsection.\arabic{subsubsubsection}}

\newcommand\subsubsubsection[1]{
    \par\refstepcounter{subsubsubsection}
    \noindent\textbf{\thesubsubsubsection. #1}
    \par}

\usepackage{amssymb}

\journal{ArXiv}

\begin{document}

\begin{frontmatter}



\title{A PRISMA-driven systematic mapping study on system assurance weakeners}


\author[york]{Kimya Khakzad Shahandashti}
\author[york]{Alvine B. Belle}
\author[uottawa]{Timothy C. Lethbridge}
\author[york]{Oluwafemi Odu}
\author[york]{Mithila Sivakumar}


            \affiliation[york]{organization={Lassonde School Of Engineering, York University},
            city={Toronto},
            country={Canada}}
\affiliation[uottawa]{organization={Department of Electrical Engineering and Computer Science, University of Ottawa},
            city={Ottawa},
            country={Canada}}

\begin{abstract}
\textbf{\textit{Context}}: An assurance case is a structured hierarchy of claims aiming at demonstrating that a given mission-critical system supports specific requirements (e.g., safety, security, privacy). The presence of assurance weakeners (i.e., assurance deficits,
logical fallacies) in assurance cases reflects insufficient evidence, knowledge, or gaps in reasoning. These weakeners can undermine confidence in assurance arguments, potentially hindering the verification of mission-critical system capabilities. This could result in catastrophic outcomes (e.g., loss of lives). It is therefore essential to
research effective solutions for managing assurance weakeners.

\textbf{\textit{Objectives}}: As a stepping stone for future research on assurance weakeners, we aim to initiate the first comprehensive systematic mapping study on this subject.

\textbf{\textit{Methods}}: We followed the well-established PRISMA 2020 and SEGRESS guidelines to conduct our systematic mapping study. We searched for primary studies in five digital libraries and focused on the 2012-2023 publication year range. Our selection criteria focused on studies addressing assurance weakeners at the modeling level, resulting in the inclusion of 39 primary studies in our systematic review.

\textbf{\textit{Results}}: Our systematic mapping study reports a taxonomy (map) that provides a uniform categorization of assurance weakeners and approaches proposed to manage them at the modeling level. The taxonomy classifies weakeners in four broad categories of uncertainty: aleatory, epistemic, ontological, and argumentation uncertainty. Additionally, it classifies approaches supporting the management of weakeners in three
main categories: representation, identification and mitigation approaches. 

\textbf{\textit{Conclusion}}:  Our study findings suggest that the SACM (Structured Assurance Case Metamodel) —a standard specified by the OMG (Object Management
Group) — may be the best specification to capture structured arguments and reason
about their potential assurance weakeners. This is due to SACM interoperability and expressiveness compared with other notations (e.g., GSN, CAE). Our findings also suggest novel assurance weakener management approaches should be proposed to assure mission-critical systems better.

\end{abstract}

\begin{keyword}
Assurance cases, Assurance deficits, Uncertainty, Logical fallacies, PRISMA, GSN, SACM, systematic mapping study, Cyber-physical Systems, Safety, Reliability.



\end{keyword}

\end{frontmatter}


\section{Introduction}
\subsection{Opening}

Cyber-physical systems (CPSs) play pivotal roles in safety- or mission-critical applications, demanding rigorous assurance of their safe operation \cite{hartsell2021automated}. Many CPSs, such as autonomous vehicles and unmanned aerial vehicles, harness the power of Machine Learning. However, these systems sometimes struggle with accurate pattern recognition and prediction when confronted with unanticipated edge cases \cite{belle2023position}. Given their multifaceted functionalities, these intricate systems can execute crucial operations across various domains, from healthcare to aviation. The immense responsibility these systems bear, particularly in scenarios directly influencing human safety, has intensified the emphasis on their dependability \cite{liu2012safety,langari2013safety,viger2021lean,hawkins2015weaving,belle2023position, zeroual2023constructing}. Addressing this imperative, the field of system assurance has risen to the fore, leveraging assurance cases to provide evidence and articulate arguments verifying a system's adherence to its set specifications and aptitude for its designated role \cite{hartsell2021automated}. Establishing comprehensive industry standards and ensuring that autonomous technology makers adhere to them is vital for building consumer trust. Increasingly, the producers of these CPSs are using assurance cases to show regulatory bodies their adherence to current standards (e.g., ISO 26262 \cite{ramakrishna2022automating}).

Amidst the surge of attention toward the assurance process, it is essential to recognize the potential pitfalls that can undermine the reliability of assurance cases. One significant aspect that needs to be addressed is the presence of assurance \textit{"weakeners" }(i.e. assurance deficits \cite{hawkins2011new} and logical fallacies \cite{greenwell2006taxonomy}) within the assurance argumentation structure. The presence of fallacies in assurance cases can lead to false confidence in the resulting argumentation structure. A fallacious assurance case may, therefore, be inadequate to properly verify the correctness of the implementation of the requirements of the system under analysis. Hence, such a system can be prone to system failure. This can have severe consequences such as death of people and financial losses \cite{viger2021lean,lin2018support}. For example, automotive software failures have led to the loss of human lives and required automotive companies to recall their vehicles at significant costs to them \cite{washingtonpost2023tesla,menghi2023assurance}. Devising solutions to effectively manage (e.g., represent, detect, mitigate) assurance weakeners is therefore critical to yield more reliable systems. 

Several approaches have been proposed to deal with assurance weakeners. They usually focus on quantifying these assurance weakeners by relying on various assessment techniques such as Baconian probabilities, Josang’s Opinion Triangle, the Dempster-Schafer theory, and the Bayesian analysis \cite{graydon2017investigation, chechik2019software}. By assigning and propagating confidence measures throughout the argument structure, these approaches quantitatively reason about the trust in the argumentation structure \cite{chechik2019software}. However, these approaches
are not specifically geared toward addressing assurance weakeners (e.g., uncertainty) in the argumentation structure \cite{chechik2019software, maksimov2019survey}. However, addressing these weakeners from a qualitative perspective, specifically at the modelling level, allows reasoning on them at a higher level abstraction and may facilitate the elicitation of more efficient approaches to address them. Still, only a few approaches address assurance weakeners at the modelling level. It is, therefore, crucial to shed light on the under-explored but promising area of assurance weakeners modelling. 

Our systematic review (systematic mapping study) reports a taxonomy that categorizes assurance weakeners and approaches proposed to manage them at the modelling level. By utilizing this taxonomy, researchers and practitioners (e.g., corporate safety analysts and regulatory authorities) can therefore: 1) enhance their understanding of existing assurance weakeners solutions and 2) have a starting point to devise effective solutions to manage assurance weakeners, ultimately leading to more reliable mission-critical systems.

\subsection{Rationale}
In order to establish the rationale for conducting our systematic review, we first searched for related reviews that specifically addressed the concept of assurance weakeners. We summarize these related reviews in the remainder of this section. Table \ref{tab:reviews} specify the focus of each corresponding review compared to our work.

\begin{table}[htbp]
\centering
\begin{tabularx}{\textwidth}{p{2.5cm}|p{2.5cm}|p{3cm}|X}
\hline
 \textbf{Reference} & 
\textbf{Publication Year} &
 \textbf{Focus} & \textbf{Title}  \\
\hline
Greenwell et al. (2005) \cite{holloway2005taxonomy}& 2005 & Classification of Logical Fallacies & A Taxonomy of Fallacies in System Safety Arguments \\
\hline
Greenwell et al. (2006) \cite{greenwell2006taxonomy} & 2006 & Classification of Logical Fallacies & A Taxonomy of Fallacies in System Safety Arguments \\
\hline
Ramirez et al. \cite{ramirez2012taxonomy} & 2012 & Classification of Uncertainty & A Taxonomy of Uncertainty for Dynamically Adaptive Systems \\
\hline
Duan et al. \cite{duan2017reasoning} & 2017 & Classification and Assessment of Uncertainty & Reasoning About Confidence and Uncertainty in Assurance Cases: A Survey \\
\hline
Graydon and Holloway \cite{graydon2017investigation} & 2017 & Confidence Assessment & An Investigation of Proposed Techniques for Quantifying Confidence in Assurance Arguments  \\
\hline
Maksimov et al. \cite{maksimov2019survey} & 2019 & Assessment of structure and content & A Survey of Tool-Supported Assurance Case Assessment Techniques  \\
\hline
Mohamad et al. \cite{mohamad2021security} & 2021 & Security Assurance Cases & Security Assurance Cases - State of the Art of an Emerging Approach \\
\hline
Khakzad et al. (our survey)& 2023 & Assurance Weakeners & A PRISMA-driven Systematic Review on Assurance Weakeners \\
\hline
\end{tabularx}
\caption{Comparison with Related Reviews}
\label{tab:reviews}
\end{table}

Greenwell et al. \cite{holloway2005taxonomy} introduced a classification system for logical fallacies found within safety cases. The purpose of this system was to provide a framework for developers and regulators, enabling them to pinpoint and rectify these logical fallacies. They grouped these into categories of relevance, acceptability, and sufficiency fallacies. A subsequent refinement by Greenwell et al. \cite{greenwell2006taxonomy} expanded on this, outlining a taxonomy that comprised 33 typical logical fallacies in safety arguments. These fallacies were grouped into seven distinct categories: Circular Reasoning, Diversionary Arguments, Fallacious Appeals, Mathematical Fallacies, Unsupported Assertions, Omission of Key Evidence, and Linguistic Fallacies.

Ramirez et al. \cite{ramirez2012taxonomy} shifted their gaze to uncertainties within Dynamically Adaptive Systems (DAS), especially in the context of intelligent vehicle systems. They crafted a taxonomy capturing potential sources of uncertainty throughout system development and spotlighted methods to tackle particular uncertainties. Contrarily, Duan et al. \cite{duan2017reasoning} categorized uncertainty as aleatory and epistemic and examined how these types were tackled in assurance cases. These studies predominantly emphasized uncertainty. Graydon and Holloway \cite{graydon2017investigation}, on the other hand, delved into methodologies for gauging confidence in assurance arguments. In a similar vein, Maksimov et al. \cite{maksimov2019survey} undertook an analysis of how assurance cases (specifically measuring confidence and uncertainty) are evaluated within various software tools. Nevertheless, their primary focus remained on the evaluative dimensions of assurance cases. Mohamad et al. \cite{mohamad2021security} embarked on a thorough review dedicated to Security Assurance Cases (SAC), a methodological approach to scrutinizing system security parameters. Their survey was notably centred around security assurance cases, bypassing other forms such as safety cases.

Therefore, the analysis of existing reviews indicates that none of them have proposed a comprehensive classification of the various assurance weakeners or a systematic overview of approaches proposed to address them. Our work, in contrast, aims to bridge this gap by focusing on the representation, detection, and mitigation of various types of assurance weakeners, including aleatory uncertainty, epistemic uncertainty, ontological uncertainty, and argument uncertainty (logical fallacies), as well as doubts (logic and epistemic doubt). Furthermore, our systematic review exclusively focuses on approaches addressing assurance weakeners at the model level. Addressing assurance weakeners at such a high level of abstraction enables more comprehensive and abstract reasoning about these weakeners, thereby simplifying their management. Our systematic review is, therefore, poised to significantly contribute to filling this crucial gap in the existing literature.

\subsection{Research Questions}
The goal of our systematic mapping study is to answer the following research questions (RQs):

\textbf{RQ1: What are the trends, patterns and relationships characterizing the literature on assurance {\textit{"weakeners"}}}? This research question aims to provide some descriptive statistics to identify trends, patterns and relationships among the primary studies, authors, venues, and related information we retrieved from the primary studies. This will notably highlight the thematic evolution of the literature on assurance weakeners and help identify the most influential themes characterizing that literature.

\textbf{RQ2: What are the different assurance \textit{"weakeners"} categories?    
 }
 This research question aims to classify various forms of \textit{"assurance weakeners"} found in the assurance case literature. That literature usually refers to assurance \textit{"weakeners"} using several keywords: assurance deficits, uncertainty, fallacies,  doubts, etc. We aim to unify that vocabulary by creating a map that structures it. This will provide future work with a controlled, unambiguous, and consistent vocabulary that can be used to refer to assurance weakeners.

\textbf{RQ3: Which approaches allow managing assurance \textit{"weakeners"}?
    } The research question at hand delves into the realm of managing assurance weakeners, seeking to identify diverse approaches that can effectively handle these weakeners. Hence, the study aims to examine the approaches the literature has proposed to manage (e.g., represent, detect, fix/mitigate) assurance weakeners. When it comes to approaches used to represent assurance weakeners, the study focuses on the various notations (e.g. graphical, textual) that the literature used to represent assurance "weakeners". In this regard, our study will notably classify (map) the different formalisms (e.g. metamodels, languages) that are usually used to formalize assurance weakeners and the associated concepts (e.g., counter-evidence, counterclaims, and counter-arguments)/structures (e.g., confidence maps) and relationships. This research question will allow us to determine whether existing assurance case notations need to be extended to better support managing the complete range of assurance weakeners we discuss in this paper.

To answer (RQ1), we perform a bibliometric synthesis \cite{linnenluecke2020conducting}. We perform a qualitative synthesis to answer (RQ2) and (RQ3). Section \ref{bibliometricS} and \ref{qualitativeS}, respectively, report both syntheses.

\section{Background Concepts}
\subsection{What is System Assurance, and how to support it?}
Ensuring the reliability and trustworthiness of software is a fundamental component of the software development process, aiming to reduce potential hazards and guarantee superior results \cite{mansourov2010system}. A method to achieve system assurance is by employing assurance cases. As depicted by Mansourov and Campara \cite{mansourov2010system}, an assurance case is characterized as a \textit{"collection of verifiable claims, arguments, and evidence assembled to confirm that a particular system/service aligns with given requirements”}. An assurance case is a document that streamlines communication among various system stakeholders (e.g., suppliers, acquirers), as well as between the operator and regulator. Its main function is to effectively convey insights about the system's prerequisites, such as safety, security, and reliability \cite{belle2023evidence,hawkins2015weaving,health2012evidence}. There exist different classifications of assurance cases, each focusing on attributes, including those centred on dependability, safety, security, and specifically, reliability and trustworthiness \cite{yuan2016automatically,jarzkebowicz2020representing,bloomfield2007confidence,rushby2013logic,cioroaica2022toward}.

The implementation of assurance cases has gained popularity, especially in sectors where safety is paramount, like healthcare, railways, automotive, and aviation \cite{duan2017reasoning,maksimov2019survey}. As a result, creating compelling assurance cases often becomes essential to receive regulatory endorsement for these systems \cite{nair2013classification}. This approach aids in verifying the safety, security, reliability and trustworthiness of mission-critical systems, mitigating severe repercussions like life loss, major injuries, environmental threats, property damages, and financial losses \cite{langari2013safety,de2016industrial}. Notably, various industry standards, such as DO-178C (in the avionics field)\cite{johnson1998178b} and ISO 26262 (pertaining to automotive), advocate for assurance cases.

\subsection{How to Represent an Assurance Case?}

Assurance cases can be represented using unstructured text (natural language) \cite{hawkins2015weaving}, semi-structured text, and graphical notations. Graphical notations have gained popularity due to their ability to express clear and well-structured arguments \cite{wei2019model}. Graphical notations include the GSN (Goal Structuring Notation) \cite{gsn} and CAE (Claims-Arguments-Evidence) \cite{bishop2000methodology}.  GSN is currently the most widely used graphical notation for representing assurance cases \cite{kelly2004goal, strunk2006essential}. The GSN Standard Working Group specifies the GSN \cite{gsn}. To enhance standardization and interoperability, the Object Management Group (OMG) recently introduced the Structured Assurance Case Metamodel (SACM) for assurance case representation \cite{omg2013sacm}. Both GSN and CAE are aligned with SACM. The nature of SACM arguments varies: they can be boolean, probabilistic,  qualitative, etc \cite{vierhauser2019interlocking}. Probabilistic arguments can be based on simulations, while qualitative arguments may be based on regulations or even the specifications of a manufacturer \cite{vierhauser2019interlocking}.

\begin{figure}[t]
  \centering
  \includegraphics[width=0.8\textwidth]{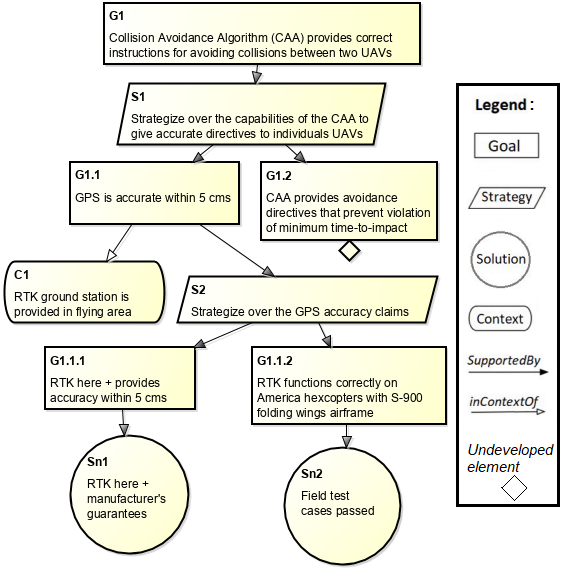}
  \caption{Partial safety case for UAV Collision Avoidance (adapted from \cite{vierhauser2019interlocking})}
  \label{fig:gsn}
\end{figure}

GSN enables the representation of an assurance case through a tree-like structure known as a \textit{goal structure}. Its core elements are goals, solutions, strategies, assumptions, contexts, and justifications \cite{gsn}. Goals in GSN represent claims, strategies depict arguments with inference rules to derive claims from sub-claims, and solutions represent evidence. Assumptions entail suppositions about claims, and justifications explain the rationale behind inferences or claims, and contexts provide contextual information for claim interpretation. GSN elements are interconnected using two types of relationships: \textit{SupportedBy} and \textit{InContextOf}. \textit{SupportedBy} allows specifying inferential or evidential relationships. An arrow with a black head represents it.  \textit{InContextOf} indicates contextual relationships. An arrow with a white head depicts it \cite{gsn}. Two decorators allow decorating GSN elements: \textit{uninstantiated} and \textit{undeveloped}. They respectively allow specifying that a GSN element (e.g., a goal) has not been instantiated yet or has not been developed and instantiated yet. Figure \ref{fig:gsn} provides an example of a safety assurance case in the GSN developed for UAV (Unmanned Aerial Vehicle) Collision Avoidance \cite{vierhauser2019interlocking}.

\subsection{The Notion of Assurance Weakeners}
In the literature on assurance cases, several expressions have been used to refer to \textit{assurance weakeners}. 
For instance, an \textit{assurance deficit} refers to \textit{"a gap in knowledge that prohibits us from having complete confidence in the assurance case"} \cite{hawkins2011new}. Additionally, according to Goodenough et al. \cite{goodenough2013eliminative}, arguments about system properties are open to revision based on new information. In the realm of \textit{defeasible reasoning}, these challenges are termed \textit{defeaters}. Hence, \textit{defeaters} are elements that challenge the validity of an argument \cite{graydon2014towards}. \textit{Defeaters} usually come in three forms \cite{jarzbowicz2015integrating, goodenough2013eliminative}: \textit{rebutting}, which provides a counter-example; \textit{undermining}, which raises doubts about evidence; and \textit{undercutting}, which questions conclusions under specific conditions when initial premises are true. 

In \cite{rushby2013logic,rushby2014mechanized}, the notion of assurance weakeners takes the form of the \textit{doubts} that may undermine the trust in the assurance of a system:   \textit{epistemic doubt} and  \textit{logic doubt}. \textit{Epistemic doubt} refers to the completeness and accuracy of the knowledge regarding the various aspects of the system at hand. These aspects include the knowledge about its requirements, environment, implementation, and hazards. \textit{Logic doubt} refers to the accuracy of the reasoning about the system's design at hand, given the knowledge we have about the system.

Burgueño et al. \cite{burgueno2019belief} defines \textit{uncertainty} as \textit{"the presence of imperfect and/or unknown information"}. That uncertainty applies to a system's predictions, estimations, measurements, and properties. According to 
Gansch et al. \cite{gansch2020system} as well as Schleiss et al. \cite{schleiss2022towards}, there are {three} types of uncertainty: \textit{aleatory uncertainty}, which means that it stems from inherent variation, \textit{epistemic uncertainty}, which means that it arises from incomplete knowledge during the modelling process and \textit{ontological uncertainty} which means as a condition of complete ignorance in the model of a relevant aspect of the system. 

Muram et al. \cite{muram2018preventing} highlight that creating arguments is complex and error-prone. It mentions that an argumentation fallacy is \textit{"a mistake or flaw in the reasoning of an argument"}. Such fallacies may lead to overconfidence in a system and the acceptance of certain faults, contributing to safety-related failures. In safety arguments, fallacies come in various forms. Greenwell et al. \cite{greenwell2006taxonomy} categorized common \textit{logical fallacies} into three types: \textit{relevance fallacies}, which offer no meaningful evidence; \textit{acceptability fallacies}, which provide contradictory or insufficient evidence; and \textit{sufficiency fallacies}, which illustrate a lack of enough evidence to support claims.

Interestingly, studies such as \cite{jarzbowicz2015integrating, jarzkebowicz2020representing} use aforementioned terms interchangeably, including assurance deficits, defeaters, and rebuttals. For consistency's sake, we refer to these various terms as \textit{ \textbf{assurance weakeners}} since they usually undermine the confidence one may have in the assurance argumentation structure.

\section{Methodology}
To carry out our systematic mapping study, we followed: 1) the guidelines for reporting systematic reviews that PRISMA (Preferred Reporting Items for Systematic Reviews and Meta-Analysis) 2020 \cite{page2021prisma}; 2) the guidelines that SEGRESS (Preferred Reporting Items for Systematic Reviews and Meta-Analysis) \cite{kitchenham2022segress}; and 3) the guidelines for reporting systematic mapping studies that Petersen et al.  \cite{petersen2015guidelines} propose. We further describe our methodology below.

\subsection{Information Sources}
In the following section, the information sources, namely database sources and snowballing sources, will be discussed.

\subsubsection{Database Sources}
To conduct the survey, we utilized five common academic databases, namely Scopus \cite{scopus}, Google Scholar \cite{googlescholar}, IEEE Xplore \cite{ieee}, ACM \cite{acm} and Engineering Village \cite{engineeringvillage} to gather relevant papers. To automatically search studies in Google Scholar, we used the \textit{Publish or Perish} tool \cite{publish}. 
To manage the studies we collected from different databases, we utilized a reference management tool called EndNote \cite{endnote}.

\subsubsection{Snowballing Sources}
We complement the database-driven search by relying on backward and forward snowballing. Wohlin \cite{wohlin2014guidelines} describes in detail how to perform snowballing. We rely on a well-established tool called \textit{Connected Papers}  \cite{connectedpapers} to automate the snowballing process. Our snowballing process uses as a start set the primary studies found through the database-driven search.

\subsection{Identification Strategies}
This section discusses the identification strategies for both database-driven search and snowballing.

\subsubsection{Database-driven Search}
In order to increase the likelihood of capturing all relevant studies in the field, it is essential that the search string employed in the search engines incorporates commonly used keywords found in papers within this particular domain. Consequently, prior to constructing the search string, we thoroughly acquainted ourselves with the precise terminology commonly utilized by researchers in the field of assurance cases. We also refined the search string multiple times until we found a search string that captured a meaningful number of studies focusing on the surveyed topic.

Next, we created the search string for the mentioned databases to identify potentially relevant studies to this study.
For that purpose, we used two groups of keywords. The first group relates to assurance weakeners, while the second one relates to assurance cases. Consequently, we formed the search string as the following:
\newline
\newline
\fbox{\begin{minipage}{\textwidth}
\textbf{
("assurance deficits" OR “false assurance” OR "defeaters" OR "counter evidence" OR "counter-argument" OR "fallacies" OR "aleatory uncertainty" OR "aleatoric uncertainty" OR "epistemic uncertainty" OR "uncertainty reasoning" OR “uncertainty elicitation” OR “informal semantics” OR “doubt”) 
\newline
AND 
\newline
(“assurance case” OR “safety case” OR “trust case” OR “dependability case” OR “reliability case” OR “security case” OR “availability case”)
}
\end{minipage}}

Due to the variations in syntax among different databases for queries, we adjusted the query for each specific database. Table \ref{tab:queries} in Section \ref{queries} detail the queries and parameters employed for the search across the five databases.

\subsubsection{Snowballing} 
We leveraged the Connected Papers tool to execute the snowballing process automatically. We iteratively relied on Connected Papers to perform the snowballing process, using the primary studies found through the database-driven search as a start set. We stopped our snowballing iterations when no additional studies were found.

\subsection{Selection Strategy}

\subsubsection{Eligibility Criteria}
To select our primary studies, we relied on a set of inclusion and exclusion criteria. Table \ref{tab:eligibility-criteria} reports these eligibility criteria. We set several criteria for inclusion to focus on papers that align with our research objectives. Firstly, we only considered papers published in English to maintain uniformity and ease of analysis, as it is the common language among the authors. Additionally, we restricted the inclusion of peer-reviewed journals and conferences. To maintain relevance and currency, we decided to focus on papers published within the last decade, i.e. from January 2012 to October 2023. Moreover, studies focusing primarily on uncertainty assessment without reasoning about uncertainty at the modeling level were excluded from our eligibility criteria. The rationale is that such techniques do not specifically address assurance weakeners in argumentation structures, whereas addressing these weakeners at the modeling level offers a higher level of abstraction that could lead to more efficient solutions.


\begin{table}[ht]
\centering
\caption{Eligibility Criteria}
\label{tab:eligibility-criteria}
\begin{tabular}{p{4cm}|p{9cm}}
\hline
\textbf{Eligibility Criteria} & \textbf{Description} \\
\hline
Inclusion & The paper should be in English. \\
\cline{2-2}
& Published in a Journal, Conference, Workshop, or Magazine. \\
\cline{2-2}
& Within ten years (January 2012 - October 2023). \\
\cline{2-2}
& Papers addressing assurance deficits and logical fallacies when creating, designing, representing, reviewing, or arguing in the assurance case. \\
\hline
Exclusion & Books, Tutorials, and Book Chapters. \\
\cline{2-2}
& Papers not available or accessible. \\
\cline{2-2}
& Serial publications. \\
\cline{2-2}
& Papers that do not discuss techniques to deal with assurance deficits or uncertainty. \\
\cline{2-2}
& Secondary Studies (e.g., Surveys, Reviews, Systematic Reviews, Systematic Literature Review, Systematic Mapping Studies), and Tertiary Studies. \\
\cline{2-2}
& Papers focusing on uncertainty assessment without reasoning about uncertainty at the modeling level. \\
\cline{2-2}
& Papers that have not been peer-reviewed. \\
\hline
\end{tabular}
\end{table}

\subsubsection{Database-driven Search}
\label{databaseS}

Table \ref{tab:selection-strategy} outlines the selection strategy we used to narrow down the studies for inclusion in the search through the database-driven search. That selection strategy consists of 6 rounds to filter the studies that did not meet the specified eligibility criteria. 


\begin{table}[htbp]
\centering
\caption{Selection Strategy}
\label{tab:selection-strategy}
\begin{tabularx}{\textwidth}{|p{2cm}|X|p{3cm}}
\hline
\textbf{Selection Round} & \textbf{Round for the Selection of Primary Studies} \\
\hline
1 & Importing in EndNote the references of studies found in the searched databases\\
\hline
2 & Cleaning references (e.g., with no title, of study type not covered, not peer-reviewed)  \\
\hline
3 & Removing duplicates (i.e., identical references coming from different databases) \\
\hline
4 & Excluding studies based on the titles, keywords, abstracts, venues, and inclusion and exclusion criteria \\
\hline
5 & Excluding studies based on the introductions and conclusion, and inclusion and exclusion criteria \\
\hline
6 & Excluding studies based on full-text reading and inclusion and exclusion criteria \\
\hline
\end{tabularx}
\end{table}

\subsubsection{Snowballing}
\label{snowballingS}
 To ensure the completeness of our search process, we use the Connected Papers tool to complete snowballing. The latter follows a similar process as the database-driven search, with slight modifications in the starting set and duplicate removal steps. Thus, in the first snowballing Iteration, we generate a graph for each primary study found through the database-driven search. Each graph generated by Connected Papers usually yields 41 studies. We then apply inclusion and exclusion criteria to each of these studies found using Connected Papers. This can result on a set of additional primary studies. We then iteratively use Connected Papers to apply snowballing on that set until no additional primary study is found.

\subsection{Data Extraction Strategies}
\subsubsection{Data Extraction Strategy for the Bibliometric Synthesis}

We reviewed the primary studies and compiled a CSV file using Notion \cite{notion2023} to document data such as publication year, venue, and number of citations, which were essential for our bibliometric synthesis. We sourced the citation count by manually exploring Google Scholar. We exported the primary studies from EndNote in RIS format to obtain data on influential research topics.

\subsubsection{Data Extraction Strategy for the Qualitative Synthesis}
We iterated over the primary studies and used research questions RQ2 and RQ3 to identify categories that could help us classify the qualitative information retrieved from these studies in accordance with both research questions. This allowed us to obtain two broad categories we further discuss in our qualitative synthesis (Section \ref{qualitativeS}). Table \ref{tab:categories}, maps each research question to its respective categories. We used these categories to create data extraction forms in Notion. We then used these forms to extract data from the primary studies. 

\begin{table}[htbp]
  \centering
  \caption{Data Extraction Categories}
  \label{tab:categories}
  \begin{tabular}{p{0.2\linewidth}|p{0.8\linewidth}}
    \hline
    \textbf{Research Questions} & \textbf{Associated Categories}\\
    \hline
    RQ2 & Categories of Uncertainty (e.g. aleatory uncertainty)  \\
    \hline
    RQ3 & Categories of approaches proposed to manage (e.g., represent, identify/detect and mitigate) assurance weakeners\\
    \hline
  \end{tabular}
\end{table}
Two researchers extracted data from the primary studies. The first researcher (a graduate student) extracted all the studies' data. The second researcher (a faculty member) randomly extracted data from 20 primary studies (half of the primary studies).
We held meetings to resolve potential disagreements between researchers during the data extraction process.

\subsection{Synthesis Strategies}
\subsubsection{Strategy for the Bibliometric Synthesis}

The bibliometric synthesis is becoming increasingly popular in several research areas \cite{linnenluecke2020conducting, neto2022safety, catumba2023sustainability, deng2023state, khanra2020big}. We, therefore, decided to use it to synthesize the various trends and patterns characterizing primary studies. To extract and synthesize the bibliometric data found in primary studies, we mainly use a popular bibliometric tool called VosViewer \cite{vosviewer}. That tool allows us to automatically generate charts depicting the bibliometric information characterizing the primary studies. To generate additional charts, we employed Python, specifically using the Pandas library \cite{pandas} for data manipulation and Matplotlib \cite{matplotlib} for data visualization. Python is an industry-standard data analysis tool with robust libraries like Pandas. 

\subsubsection{Strategy for the Qualitative Synthesis}
To perform our qualitative synthesis, we have used tables to capture and report the data extracted from primary studies. This is useful to present data in a tabular form in accordance with the research questions associated with the qualitative synthesis. Presenting data in a tabular form is a practice that is strongly recommended when completing systematic reviews  \cite{kitchenham2004procedures}.

\section{Results} 
\subsection{Study Selection}
\begin{figure}[htbp]
  \centering
\includegraphics[width=1.1\textwidth]{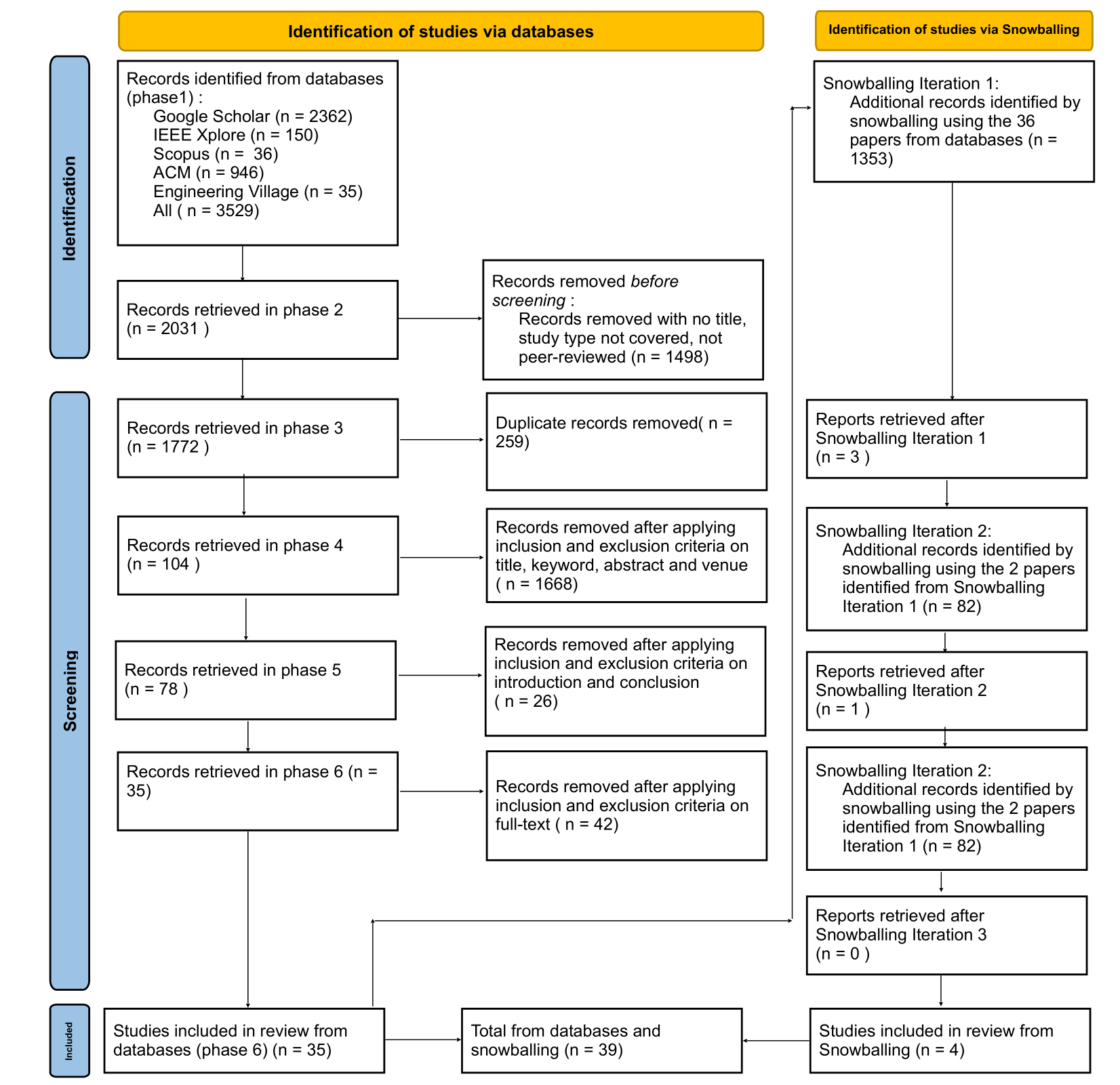}
  \caption{PRISMA flow diagram illustrating the identification and search processes}
  \label{fig:prisma_flow}
\end{figure}

Figure \ref{fig:prisma_flow} shows our PRISMA flow chart diagram. This diagram illustrates the selection results that both our database-driven search and snowballing strategies yield. This diagram maps out the number of records identified, screened and included from different information sources. 
Overall, 39 primary studies were selected from the database and snowballing search.

Noteworthy, as of October 10, 2023, Connected Papers could not snowball some publications (i.e. \cite{millet2023assurance}, \cite{diemert2023incremental}  and \cite{belle2023position}) due to their recency.  

\subsection{Study Characteristics}
\ref{studychar} reports the characteristics of the primary studies. The columns of that table provide information about the authors of the primary studies, their publication year, their title, venue, and the type of search (e.g., database-driven search, snowballing) used to select them. The venues' acronyms are also shown in \ref{acronyms}.


\subsection{Bibliometric Synthesis Results}
\label{bibliometricS}
We answer (RQ1) in this section. For this purpose, we generate charts embedding bibliometric information and comment on them. We further describe our bibliometric synthesis in the remainder of this section.

\subsubsection{Publication Year Trend}
\begin{figure}[t]
  \centering
\includegraphics[width=0.9\textwidth]{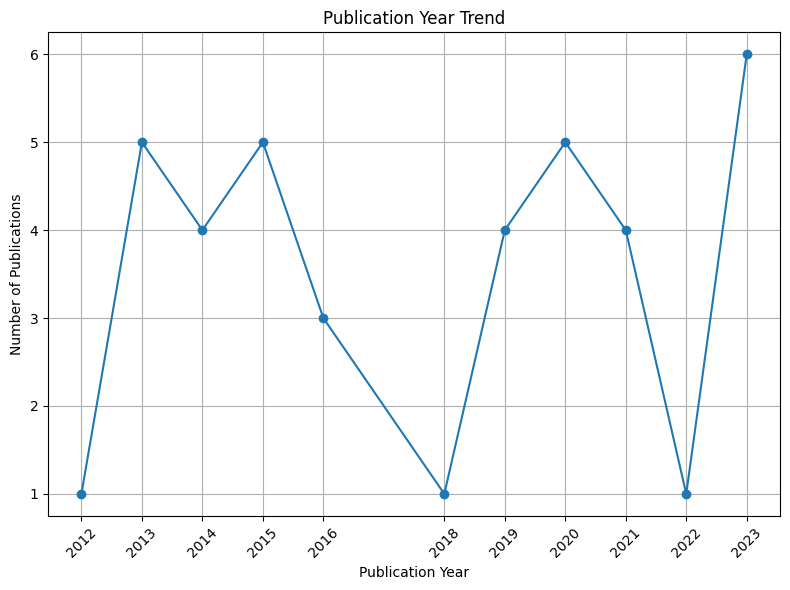}
  \caption{Publication Year Trends}
  \label{fig:trend_chart}
\end{figure}

Figure \ref{fig:trend_chart} illustrates the publication year trend for the primary studies. A notable peak of five publications in 2013 signalled increased interest in the field compared to the previous year. Subsequently, the number of studies displayed fluctuations until a notable dip to a single study in 2018. However, a subsequent recovery was observed leading up to 2020. The decline after 2020 could be attributed to various factors, including possible repercussions from the global COVID-19 pandemic. The pandemic's disruptive effects on research activities, ranging from practical limitations and resource reallocation to financial uncertainties, likely contributed to this decrease. Nonetheless, in 2023, the trend has seen an unprecedented surge. This is possibly due to the introduction of innovative methods, increased funding, or the growing topicality and cruciality of the surveyed theme, especially regarding safety-critical systems such as cyber-physical systems. This recent spike in studies publications underscores assurance weakeners' growing significance and relevance, reflecting its prominence in the research community and the industrial sector.

\subsubsection{Venue Distribution}
\begin{figure}[t]
  \centering
\includegraphics[width=0.9\textwidth]{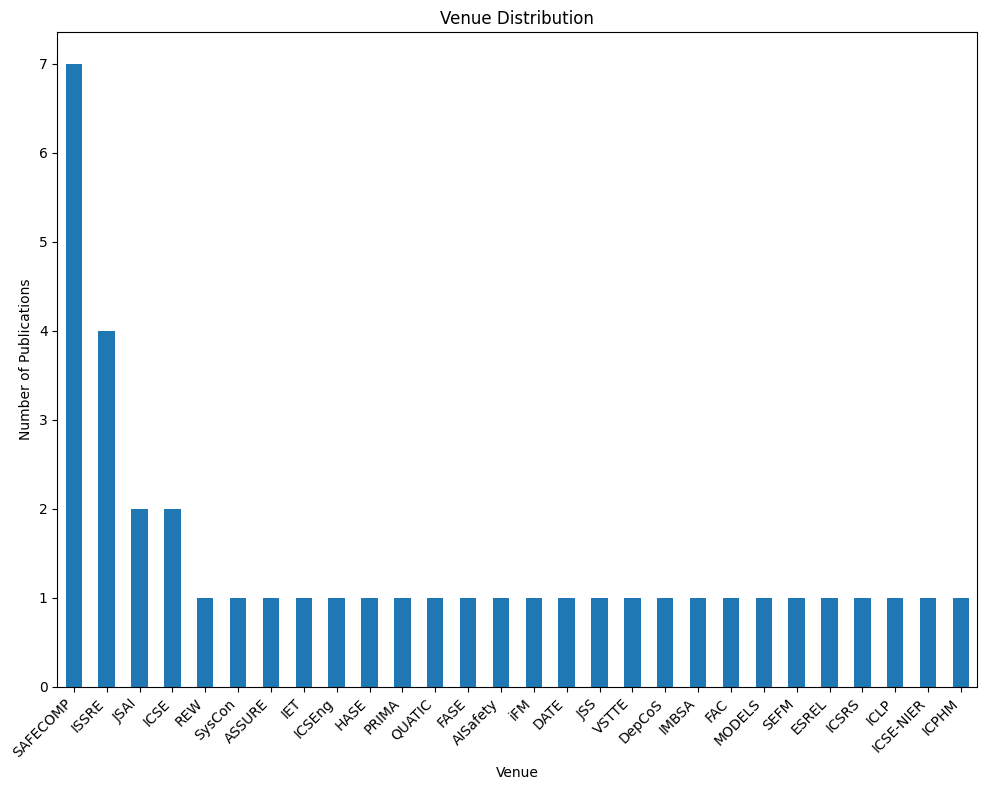}
  \caption{Venue Distribution}
\label{fig:venue_distribution}
\end{figure}

Figure \ref{fig:venue_distribution} presents a detailed visualization of the distribution of studies across various venues. The most striking observation is the dominance of SAFECOMP, which has the most publications among the venues considered. SAFECOMP's preeminence in primary studies can be attributed to its comprehensive focus on developing, assessing, operating, and maintaining safety-related and safety-critical computer systems. This wide-ranging purview harmonizes seamlessly with the core research theme on assurance weakeners. Moreover, SAFECOMP’s esteemed reputation in the fields of safety, reliability, and security further underpins its high publication count.

Next in line, the International Symposium on Software Reliability Engineering (ISSRE) has a considerable number of studies, which can be attributed to its strong alignment with the domain of system assurance. ISSRE's emphasis on software reliability engineering closely resonates with the fundamental concerns of system assurance and the mitigation of assurance weakeners.

Following ISSRE, the International Conference on Software Engineering (ICSE) also stands out with a significant concentration of studies. ICSE's prominence in the illustrated distribution can be attributed to its status as a leading conference in the software engineering domain.

\subsubsection{Most influential research topics}
 
\begin{figure}[t]
  \centering
\includegraphics[width=0.9\textwidth]{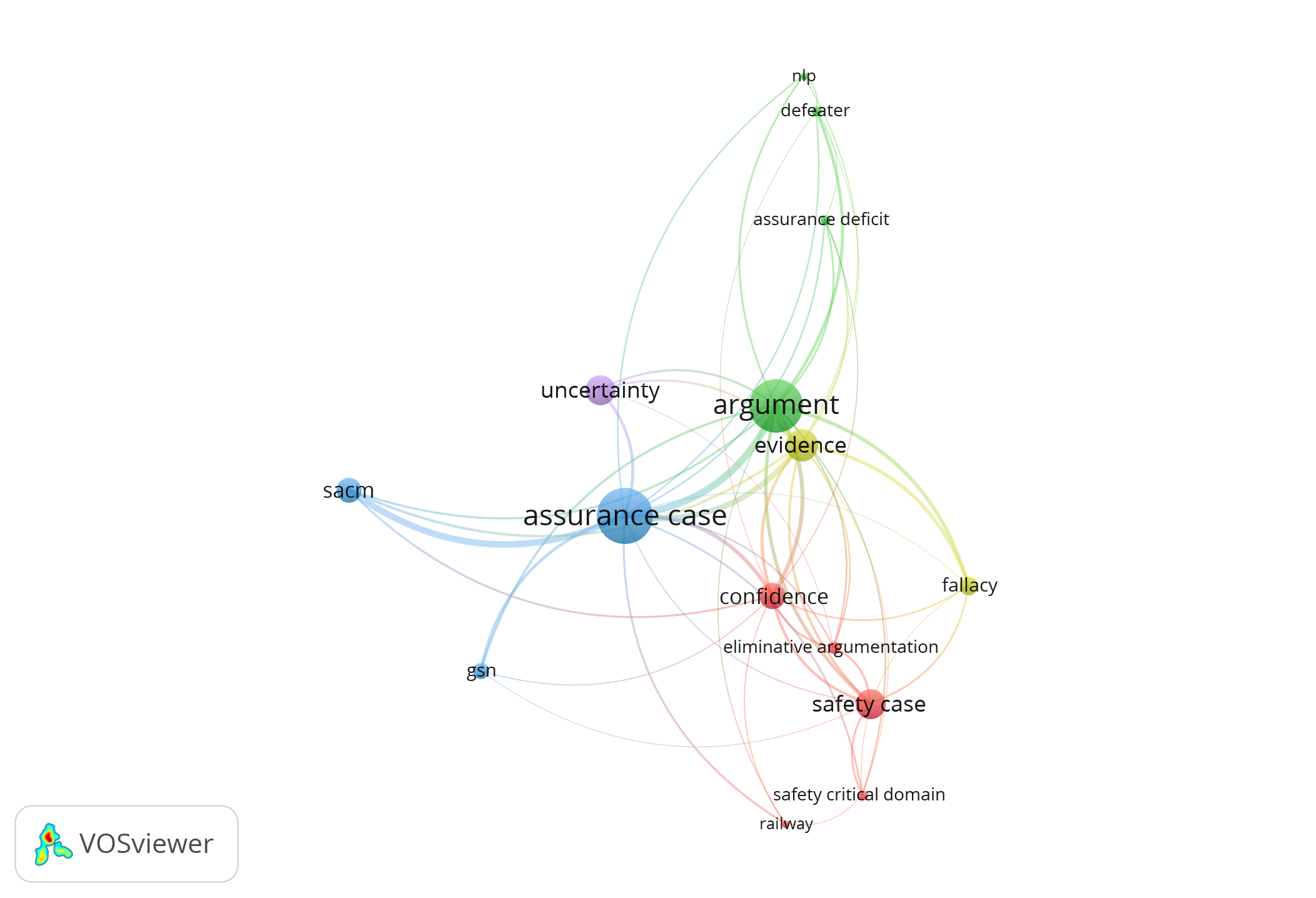}
  \caption{Most Influential Research Topics}
\label{fig:research_topics}
\end{figure}

The visual representation captured in Figure \ref{fig:research_topics} sheds light on the pivotal topics shaping the assurance weakeners domain. We generated this chart with VosViewer to create a term co-occurrence map with a minimum of 2 occurrences of a term on the title and abstracts field. We also removed the noisy keywords.

The term \textbf{"assurance case"} emerges prominently as a central pivot in the discussion, emphasizing its critical importance. This concept serves as an anchor for other interrelated terms, reflecting the intricate network of these themes. Notably, nestled close to \textbf{"assurance case"} is the \textbf{"safety case"}, which underscores the significant emphasis the research community places on safety assurance. When compared to other types of assurance cases like security or reliability cases, the prominence of \textbf{"safety case"} further highlights the supreme significance of safety within the realm of assurance weakeners. The prioritization of safety is rational, considering that system safety is crucial to prevent disastrous consequences, ranging from the loss of lives to substantial economic or environmental damages. Noteworthy, the term \textbf{"safety case"} frequently appears in the contexts of \textbf{"safety-critical domains"} such as \textbf{"railway"}. This association underscores the potentially catastrophic implications of system failures in these domains, reinforcing the mandatory requirement of a \textbf{"safety case"} for certification.
Additionally, the presence of terms such as \textbf{"assurance deficit"}, \textbf{"defeater"},\textbf{"uncertainty"} and \textbf{"fallacy"} emphasize assurance weakener-related terminology. The latter brings to light potential shortcomings or lapses in the assurance process. Their prominence underscores the importance of addressing these weakeners to achieve the highest safety standards. There's also a clear linkage between \textbf{"eliminative argumentation"} and \textbf{"confidence"}. This connection suggests that eliminative argumentation is a prevalent mitigation strategy used to tackle assurance weakeners and increase system confidence.
To add, keywords such as \textbf{"gsn"} and \textbf{"sacm"} are among the most recurring keywords in Figure \ref{fig:venue_distribution}. Both notations are among the most used to represent assurance weakeners.

\subsubsection{Most Cited References}
Table \ref{tab:top_cited} provides an overview of the 15 most frequently cited references, offering a glimpse into the scholarly works that have significantly influenced the field. Among these notable references, the contributions of Denney et al. \cite{denney2015dynamic}, Yamamoto et al. \cite{yamamoto2013evaluation}, and Goodenough et al. \cite{goodenough2013eliminative} stand out as pivotal and widely acknowledged. 

The prominence of Denney et al.'s work \cite{denney2015dynamic} can be attributed to its pioneering introduction of dynamic safety cases, an assurance case for operationalizing through-life safety assurance. Published in 2015, this seminal contribution addresses critical safety concerns and advocates proactive safety management practices. Furthermore, its early publication date may have given it sufficient time to disseminate across the research community, which significantly contributed to its significant citation count.

Yamamoto et al.'s work \cite{yamamoto2013evaluation} occupies a prominent position due to its insightful exploration of assurance case development challenges and innovative solutions through assurance case pattern methods. The contribution of Goodenough et al. \cite{goodenough2013eliminative} holds significance for its pioneering application of eliminative induction and defeasible reasoning principles. Introducing the concept of a "confidence map" allows for providing a structured approach to addressing doubts and reinforcing confidence in system property assertions.

The extensive citations of renowned authors such as Ewen Denney, Ganesh Pai, Marsha Chechik, and Patrick John Graydon may stem from their established reputation as influential researchers in the assurance case domain. As well-established figures, their work naturally becomes reference points for other scholars, further amplifying their citation counts.

{\footnotesize
\begin{table}[htbp]
  \centering
  \caption{Top 15 Most Cited Papers as of October 18, 2023}
  \label{tab:top_cited}
  \begin{tabular}{p{0.13\linewidth}|p{0.3\linewidth}|p{0.45\linewidth}|p{0.15\linewidth}}
    \hline
    \textbf{Ranking} & \textbf{Authors (Reference)} &
    \textbf{Title} & \textbf{Citations} \\
    \hline
    1 & Denney et al. (2015a)  \cite{denney2015dynamic}& Dynamic safety cases for through-life safety assurance & 99 \\
    \hline
    2 & Yamamoto et al. \cite{yamamoto2013evaluation} & An evaluation of argument patterns to reduce pitfalls of applying assurance case & 61 \\
\hline
    3 & Goodenough et al. \cite{goodenough2013eliminative} & Eliminative induction:
A basis for arguing system confidence  & 48 \\
\hline
    4 & McDermid et al. \cite{mcdermid2019towards} & Towards a framework for safety assurance
of autonomous systems & 40 \\
\hline
    5 & Rushby \cite{rushby2013logic} & Logic and epistemology in safety cases & 33 \\
    \hline
    6 & Denney and Pai. \cite{denney2013evidence} & Evidence arguments for using formal methods in software
certification  & 30 \\
\hline
    7 & Muram et al. \cite{muram2018preventing} & Preventing omission of key
evidence fallacy in process-based argumentation & 24 \\
\hline
    8 & Rushby \cite{rushby2014mechanized} & Mechanized support for assurance case argumentation & 21 \\
    \hline
    9 & Denney et al. (2015b) \cite{denney2015formal} & Formal foundations for hierarchical safety
cases & 19 \\
\hline
    10 & Chechik et al. (2019) \cite{chechik2019software} & Software assurance
in an uncertain world & 18 \\
\hline
    11 & Graydon \cite{graydon2014towards} & Towards a clearer understanding of context and its role in assurance argument confidence & 16 \\
    \hline
    12 & Nemouchi et al. \cite{nemouchi2019isabelle} & Isabelle/sacm: Computer-
assisted assurance cases with integrated formal methods & 15 \\
\hline
    13 & Gansch et al. \cite{gansch2020system} & System theoretic view on uncertainties & 15 \\
    \hline
    14 & Groza et al. \cite{groza2015formal}&  A formal approach for
identifying assurance deficits in unmanned aerial vehicle software & 13 \\
\hline
    15 & Cârlan et al. (2016b) \cite{carlan2016using} & On using results of code-level bounded
model checking in assurance cases& 11 
  \end{tabular}
\end{table}
}

\subsection{Qualitative Synthesis Results}
\label{qualitativeS}
\begin{figure}[htbp]
  \centering
\includegraphics[width=1.15\textwidth]{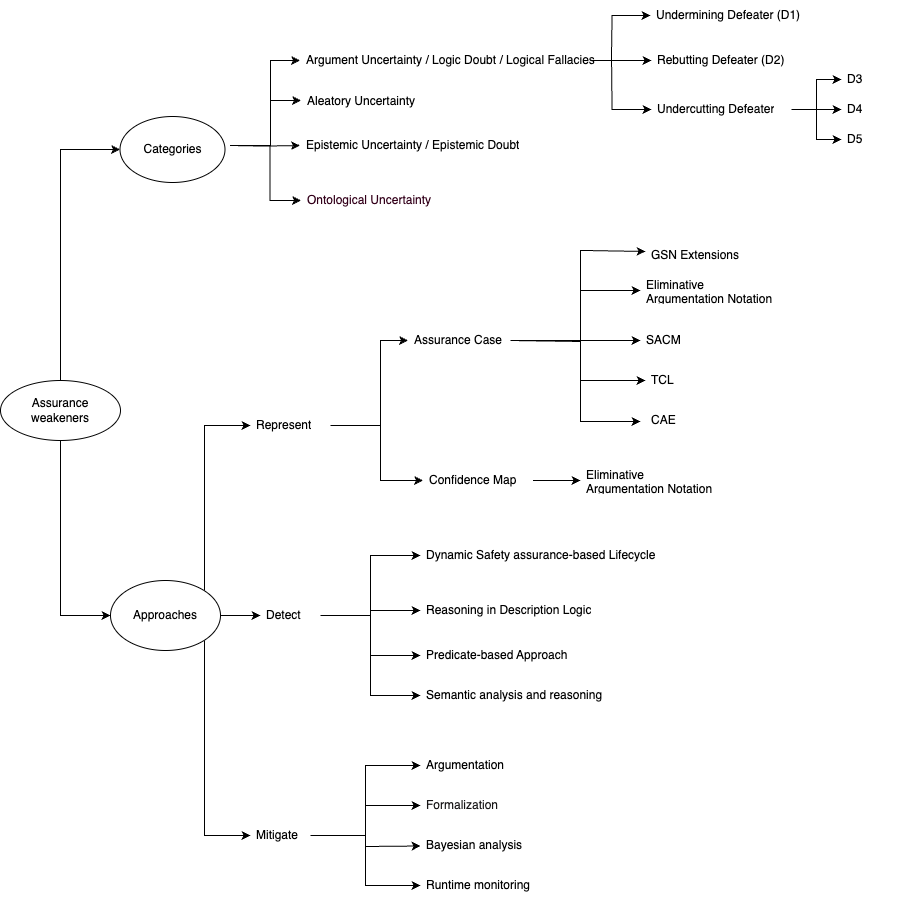}
  \caption{Taxonomy of assurance weakeners and their management approaches}
  \label{fig:taxonomy}
\end{figure}

In this section, we address both (RQ2) and (RQ3). To achieve this, we have developed Figure \ref{fig:taxonomy}, which visually represents the taxonomy that we constructed from the primary studies. This taxonomy encompasses various \textbf{categories} of assurance weakeners (RQ2) and the corresponding \textbf{approaches} proposed in the literature for their management (RQ3). This approach enables us to adapt the concepts of "types" and "means" as used by Gansch et al. \cite{gansch2020system} to investigate a specific category of weakeners, namely uncertainty, and the supporting management approaches.


\subsubsection{Categories of Assurance Weakeners (RQ2)}
\label{categorieS}

The taxonomy that Figure \ref{fig:taxonomy} depicts allows us to classify assurance weakeners into several categories of uncertainty. Uncertainty can stem from several sources: the system's specifications, the environment in which the system operates, the system itself, the extent to which the arguments about the system's capabilities can be trusted, etc. \cite{chechik2019software}.

One notable thing in our review was that some primary studies, such as the work of Yuan et al. \cite{yuan2016automatically} and Liu et al. \cite{liu2012safety}, use different classifications of logical fallacies, which pertain to evidence, claims, and arguments. Essentially, they all point to shortcomings in claims, arguments, or evidence. These types of doubts are often termed undermining, undercutting, and rebutting defeaters. Given this established terminology, we chose to use these terms to describe logical fallacies, logic doubts, and argument uncertainties. Consequently, our categorization is structured accordingly.

According to Gansch et al. \cite{gansch2020system}, Denney et al. \cite{denney2015dynamic}, Chechik et al.  \cite{chechik2019software}, and Schleiss et al. \cite{schleiss2022towards}, uncertainties can be classified into the following categories: aleatory, epistemic, ontological, and argument (logical). 

\begin{itemize}

    \item \textbf{Aleatory Uncertainty}:  it pertains to the inherent unpredictability of a particular event or situation. Such uncertainties are often described as \textit{"known unknowns"} and can be measured using probability distributions \cite{gansch2020system,denney2015dynamic,chechik2019software,schleiss2022towards,belle2023position,duan2017reasoning}.Examples of aleatory uncertainties include an overdose from an infusion pump or mistakes in computing medication dosages \cite{duan2017reasoning}. Additionally, one often-overlooked type of aleatory uncertainty is the residual kind: even when a risk is thought to be addressed, there's a slight possibility it may not have been completely mitigated \cite{duan2017reasoning}.
    \item \textbf{Epistemic Uncertainty/ Epistemic Doubt}:
    also referred to as epistemic doubt, epistemic uncertainty
    arises from a deficiency in knowledge or data, often referred to as the \textit{"unknown unknowns."}\cite{gansch2020system,denney2015dynamic,chechik2019software,schleiss2022towards,belle2023position,duan2017reasoning}. Examples encompass unnoticed flaws in logical thinking or unanticipated input sequences during the design and development phases \cite{duan2017reasoning}.

    \item \textbf{Ontological Uncertainty}: it delves into a state of complete unfamiliarity with a vital aspect of the system's representation \cite{gansch2020system,schleiss2022towards}. 
    Particularly in the initial phases of research and development for new systems, the accuracy and entirety of models often face scrutiny. For example, in autonomous systems functioning in open environments, one cannot entirely eliminate ontological uncertainty \cite{gansch2020system,bandur2015informing}.
    \item \textbf{Argument Uncertainty / Logical Fallacies / Logic Doubt:} the endeavor of constructing argumentations has been acknowledged as resource-intensive, time-consuming, and prone to errors \cite{menghi2023assurance,nair2014extended}. Inaccurate, incomplete, or inherently flawed reasoning regarding evidence can introduce defects known as logical fallacies into safety argumentations, leading to overconfidence in a system and tolerating certain faults, ultimately contributing to safety-related system failures \cite{muram2018preventing}. According to Chechik et al.  \cite{chechik2019software}, using safety arguments to convince that software-intensive systems are safe enough raises questions about the extent to which these arguments can be trusted. These questions relate to the confidence one may have in the completeness of the verification and validation processes used to support the contention that the software at hand is indeed safe. These questions embody argument uncertainties and relate to the extent to which the evidence sufficiently backs arguments and the thoroughness of the arguments. Several primary studies (i.e. \cite{sun2014rethinking,bandur2015informing,nemouchi2019isabelle}) use the term \textit{logical fallacies} to refer to argument uncertainties. A few primary studies use the expression  \textit{logic doubt} to refer to argument uncertainties (i.e. \cite{rushby2013logic,rushby2014mechanized}). 
    
    A Logic doubt is also called \textit{a defeater}. There are three main categories of defeaters: \textit{undermining}, \textit{undercutting}, and \textit{rebuttal} defeaters \cite{jarzbowicz2015integrating, goodenough2013eliminative, takai2014supplemental, grigorova2014argument,diemert2020eliminative,cobos2021cybersecurity}. When "defeaters" represent a level of uncertainty deemed permissible as remaining doubt (without additional justification), they are labelled as "residual" and add to the total lingering doubt linked to the case \cite{muram2023attest}. Jarz\k{e}bowicz and Wardzi\'{n}ski  \cite{jarzbowicz2015integrating} further divide these three categories of defeaters into five categories of defeaters, namely: D1, D2, D3, D4 and D5. We provide below the definitions of these three defeaters and explain their mapping with the five categories of defeaters that Jarz\k{e}bowicz and Wardzi\'{n}ski  \cite{jarzbowicz2015integrating} proposed. 

\begin{itemize}
    \item \textbf{Undermining Defeaters}: Invalidate one or more premises, thereby diminishing the basis for accepting the associated claim, even if the inference rule remains sound \cite{jarzbowicz2015integrating, goodenough2013eliminative, takai2014supplemental, grigorova2014argument,diemert2020eliminative,muram2023attest,cobos2021cybersecurity}. Undermining defeater is also called \textbf{D1} (i.e., provision of arguments against the evidence)\cite{jarzbowicz2015integrating}.
    
    \item \textbf{Rebuttal Defeaters}:  provide a counterexample to a claim by introducing information that contradicts them and challenge their validity \cite{jarzbowicz2015integrating, goodenough2013eliminative, takai2014supplemental, grigorova2014argument,diemert2020eliminative,cobos2021cybersecurity}. Rebutting defeater is also called \textbf{D2} (direct refutation of the claim, typically through counterevidence)\cite{jarzbowicz2015integrating}.

    \item \textbf{Undercutting Defeaters}: Focus on contesting the inference rule itself, presenting additional information about scenarios in which the claim might not hold true even with true premises \cite{jarzbowicz2015integrating, goodenough2013eliminative, takai2014supplemental, grigorova2014argument,diemert2020eliminative,muram2023attest,cobos2021cybersecurity}. The more detailed aspects of the Undercutting defeater are split into three variations \cite{jarzbowicz2015integrating}: 1) \textbf{D3} (attack on the inference rule's validity); 2) \textbf{D4}  (disruption of the link between premises and conclusion, often due to misuse of the inference rule); and 3) \textbf{D5}  (challenge to the applicability of the inference rule).
\end{itemize}
    
\end{itemize}

 \fbox{%
  \parbox{\dimexpr\linewidth-2\fboxsep-2\fboxrule\relax}{
    \textbf{Key Takeaway:}\\
   Although there are several categories of uncertainties, most of the research work centred around addressing assurance weakeners at the modeling level focuses on argument uncertainty. This may indicate that argument uncertainty may explicitly be tied to the flawed argument structure and may, therefore, be reasoned away by improving that argument structure.
  }%
}

\subsubsection{Categories of approaches proposed to deal with Assurance Weakeners (RQ3)}

The taxonomy in Figure \ref{fig:taxonomy} classifies the approaches proposed to manage assurance weakeners into the following categories:

\subsubsubsection{Approaches to Represent Assurance Weakeners} \label{repWeakenerS}
Several approaches focus on the representation of assurance weakeners. They commonly depict assurance weakeners using assurance cases or confidence maps, as shown in Figure \ref{fig:taxonomy}. Table \ref{tab:representation-approaches} reports the primary studies that focused on representing assurance weakeners using these two structures. Table \ref{tab:representation-approaches} also map these primary studies to the different notations they used when representing assurance weakeners. These notations include GSN, SACM, CAE and TCL, as well as their extensions (when applicable).

\begin{table}[htbp]
\centering
\caption{Approaches for Representing Assurance Weakeners}
\label{tab:representation-approaches}
\begin{tabular}{>{\raggedright}p{3.5cm} p{3.5cm} p{5cm}}
\toprule
\textbf{References} & \textbf{Representation structure} & \textbf{Notation} \\
\midrule
\cite{graydon2014towards,groza2015formal,denney2015formal,takai2014supplemental} & Assurance Case & GSN Extension \\
\addlinespace
\cite{millet2023assurance,menghi2023assurance,diemert2023incremental,cobos2021cybersecurity} & Assurance Case & Eliminative Argumentation \\
\addlinespace
\cite{jarzbowicz2015integrating} & Assurance Case & TCL \\
\addlinespace
\cite{selviandro2020visual,foster2021integration} & Assurance Case & SACM \\
\addlinespace
\cite{murugesan2023semantic} & Assurance Case & CAE \\
\addlinespace
\cite{goodenough2013eliminative,diemert2020eliminative} & Confidence Map &  Eliminative Argumentation \\
\bottomrule
\end{tabular}
\end{table}

\paragraph{\textbf{GSN-based representation approaches:}} The most common notation for representing assurance weakeners is the Goal Structuring Notation (GSN) \cite{chechik2019software}. However, it's important to note that GSN alone is sometimes insufficient to effectively capture weakeners. So, extending that notation is often required to properly capture assurance weakeners. GSN primarily serves as a notation for explicitly presenting assurance cases without inherently judging the quality of an argument \cite{grigorova2014argument}. Graydon \cite{graydon2014towards} advocates for assured safety arguments, encompassing two sub-arguments: one that documents the argument and evidence for system safety and another that justifies the sufficiency of confidence in this safety argument, considering plausible defeaters. Groza et al. \cite{groza2015formal} formally represent argumentative-based GSN and employ description logic for identifying assurance deficits in GSN models. Denney et al. \cite{denney2015formal} introduce hierarchical safety cases in GSN to enhance the comprehension of safety argument structures and mitigate assurance deficits, albeit noting the need for a formal specification to verify tool operations to eliminate potential deficits. In addressing uncertainty, Yuan et al. \cite{yuan2016automatically} propose a predicate-based representation of GSN elements, contributing to considering uncertainty within safety arguments. Moreover, Takai and Kido \cite{takai2014supplemental} propose a supplemental notation for GSN, rooted in defeasible logic and argumentation theory, to address unexpected changes and rebut arguments explicitly.

\paragraph{\textbf{Eliminative Argumentation-based approaches:}} the Eliminative Argumentation (EA) notation allows constructing arguments and evaluating confidence in these arguments by relying on the notion of \textit{defeasible reasoning} \cite{diemert2020eliminative}. The latter supports the recursive challenging of claims to progressively eliminate the doubts they may embed and, consequently, increase the confidence in these arguments \cite{diemert2020eliminative}.
EA notation builds on the foundational ideas of GSN. Just as in GSN, EA employs a directed, non-cyclic graph to lay out the structure of an argument. Both EA and GSN methodologies employ a hierarchical tree-like design where a principal system claim (positioned at the "root") gets broken down into more detailed sub-claims, eventually anchoring in concrete evidence. This offers a clear linkage between the evidence presented and the resulting claims. A distinctive feature of EA, as compared to GSN, is its capacity to capture and represent "doubts" or defeaters regarding the authenticity of claims, the backing evidence, or the logical conclusions derived \cite{diemert2023incremental}. Therefore, in EA, defeaters can highlight uncertainties related to claims, the adequacy of evidence in supporting its preceding claim, and the robustness or comprehensiveness of the derived conclusions. Defeaters in EA notation are represented in rectangles with chopped-off corners, and each defeater has its own color: red for rebutting, yellow for undermining, and orange for undercutting and inference rules are also represented using green rectangles \cite{goodenough2015eliminative}.
Menghi et al. \cite{menghi2023assurance}, Diemert et al. \cite{diemert2023incremental}, Millet et al. \cite{millet2023assurance} and Cobos et al. \cite{cobos2021cybersecurity} use EA notation to represent different types of defeaters.


An eliminative argument can be visualized using a \textbf{Confidence map} \cite{goodenough2015eliminative}. Accordinly, Goodenough et al. \cite{goodenough2013eliminative} rely on Confidence maps to notably represent defeaters. These maps are grounded in eliminative induction philosophy and defeasible reasoning. Confidence maps explicitly represent reasons for doubt relevant to an argument, aiding in its development and evaluation. Diemert et al. \cite{diemert2020eliminative} also employ the concept of confidence maps, providing an abstract framework for constructing and assessing arguments based on defeasible reasoning. According to Goodenough et al. \cite{goodenough2013eliminative}, a confidence map is a graphical argumentation structure that is mainly constituted from a set of claims, defeaters, inference rules, and evidence. Their work focuses on three defeaters: rebutting, undermining, and undercutting. They depict claims using clear rectangles, they depict inference rules using green-shaded rectangles, and they depict defeaters using red-shaded octagons. They label claims with a “C” and inference rules with an “IR”. They respectively label rebutting, undercutting, and undermining defeaters with an “R”, “UC,” and a “UM”. To depict the completion of the doubt refinement, they rely on the concept of argument terminators represented by a shaded circle.

\paragraph{\textbf{SACM-based representation approaches:}}
Selviandro et al. \cite{selviandro2020visual} as well as Foster et al. \cite{foster2021integration} discuss the Structured Assurance Case Metamodel (SACM), a rich specification for structured assurance cases, offering features that surpass existing notations. For instance, SACM includes the concept of "Defeated Assertion," indicating when an assertion is defeated by counter-evidence or argumentation and visualizes it as a Claim with a cross. Moreover,
a line representing a \textit{Defeated AssertedContext} is depicted with a solid square close to one end and a cross positioned at its center.
Noteworthy, V2.3 of the SACM specification  \cite{SACMv2.3}, provides more details on the various three categories of defeaters discussed in this review. Overall, with such notations and concepts, SACM therefore supports the representation of all the defeaters and their relationships. Still, very few approaches have adopted SACM so far. This implies that more effort should be made to promote SACM in the research community and in the industry.

\paragraph{\textbf{TCL-based representation approaches:}} another approach to represent is presented in Jarz\k{e}bowicz and Wardzi\'{n}ski \cite{jarzbowicz2015integrating}'s work, which is for integrating confidence and assurance arguments using the Trust Case Language (TCL) and providing a checklist of defeaters to build confidence arguments.

\paragraph{\textbf{CAE-based representation approaches:}}
Murugesan et al. \cite{murugesan2023semantic} propose a Claims-Arguments-Evidence approach, emphasizing the evidence presented, the logical reasoning utilized, and the exploration and assessment of counter-claims. Doubts, concerns, or counter-claims on any part of the case are captured as defeaters and represented as red ellipses.

 \fbox{%
  \parbox{\dimexpr\linewidth-2\fboxsep-2\fboxrule\relax}{
    \textbf{Key Takeaway:}\\
Over the years, various notations have emerged to represent assurance weakeners. However, many of these notations, while valuable in their own right, have had limitations in terms of expressiveness, adaptability, and their ability to holistically represent complex interrelationships between arguments and weakeners.
SACM, therefore stands out among existing notations. SACM is the "relatively new kid on the block", as it is a standard that unifies several notations such as GSN and CAE \cite{nemouchi2019isabelle}. SACM offers a more comprehensive range of capabilities than current system assurance notations. It lays the groundwork for model-driven system assurance, showing immense potential in that emerging specification \cite{wei2019model}.
Its superior interoperability means it can seamlessly integrate with other systems and notations, making it a preferred choice for diverse applications. Moreover, its enhanced expressiveness allows for more intricate detailing of arguments and their potential pitfalls. This adaptability ensures a comprehensive and intuitive representation of arguments, their interrelations, and any associated weakeners. Esteemed institutions like the University of York in the UK and Carnegie Mellon University have already vouched for SACM \cite{SACMv2.3}. 
Another critical takeaway is the absence of an approach representing aleatory, epistemic and ontological uncertainties, as current approaches focus on representing argument uncertainty. This is primarily because these types of assurance weakeners stem from inherent unpredictability, incomplete knowledge or unfamiliarity of the system representation(unknowns in the system). Consequently, it may be challenging to explicitly represent such types of assurance weakeners.
  
}
}\\


\subsubsubsection{Approaches to Detect Assurance Weakeners}
\label{detWeakenerS}

In this section, we explore approaches used to detect (identify) assurance weakeners, as seen in Figure \ref{fig:taxonomy}. Table \ref{tab:detection-approaches} reports the primary studies focusing on detecting assurance
weakeners. Groza et al. \cite{groza2015formal} propose using reasoning in description logic (DL) to identify assurance deficits in the GSN model. Their solution has two steps: First, they check with hybrid logic if the evidence nodes from the GSN representation have their corresponding formulas validated against the Kripke model. Second, by reasoning in DL, they identify which goals in the GSN model are not supported by verified evidence. Moreover, Denney et al. \cite{denney2015dynamic} suggest that assurance deficits (i.e. aleatory uncertainty and epistemic uncertainty) can weaken confidence in the safety and as the system and its safety argument change, so will the assurance deficits. Consequently, they claim there is a need for a dynamic safety case lifecycle to identify such deficits by proactively computing confidence and updating the reasoning about ongoing operations' safety. 

Murugesan et al. \cite{murugesan2023semantic} propose using semantic analysis to identify counter-claims (defeaters) and counter-evidence. Moreover, Muram and Javed. \cite{muram2023attest} introduces ATTEST, an assurance framework rooted in natural language processing (NLP). Initially, the framework processes text through various NLP procedures. It then formulates rules that encapsulate both syntactic attributes obtained via NLP tasks and semantic features discerned from model structures and their interconnections. These rules are subsequently activated to understand arguments, ensure their correctness, verify their adequacy, identify potential defeaters, and select counter-evidence. Regarding detecting argument fallacies, Yuan et al. \cite{yuan2016automatically}  examine a subset of them and how they can be detected automatically via predicate-based representation. They build an ontology that contains a set of constant symbols, function symbols and predicate symbols, which form the vocabulary for the expressions of GSN nodes.


\begin{table}[htbp]
\centering
\caption{Approaches for Detecting Assurance Weakeners}
\label{tab:detection-approaches}
\begin{tabular}{>{\raggedright}p{3.5cm} p{10cm} }
\toprule
\textbf{References} & \textbf{Categories of Approaches}  \\
\midrule
\cite{groza2015formal} & Reasoning in Description Logic \\
\addlinespace
\cite{denney2015dynamic} & Dynamic Safety assurance-based Lifecycle \\
\addlinespace
\cite{murugesan2023semantic,muram2023attest} & Semantic analysis and reasoning \\
\addlinespace
\cite{yuan2016automatically} & Predicate-based Approach  \\
\bottomrule
\end{tabular}
\end{table}

 \fbox{%
  \parbox{\dimexpr\linewidth-2\fboxsep-2\fboxrule\relax}{
    \textbf{Key Takeaway:}\\
 In our exploration of assurance weakener detection approaches, several methods have emerged. Predominantly, semantic analysis and reasoning, as discussed in \cite{murugesan2023semantic,muram2023attest}, have become the most utilized techniques. However, there's an evident limitation in current approaches. Many tend to concentrate on specific issues, either pinpointing argument fallacies or highlighting assurance deficits. This segmented view means we lack a comprehensive method capable of encompassing all assurance weakeners types. It underscores a pressing need for a more holistic detection approach in the field.
}
}\\

\subsubsubsection{Approaches to Mitigate Assurance Weakeners}
\label{mitWeakenerS}
Once assurance weakeners have been detected, we can decide to tolerate them (i.e. “live with them”) or to mitigate them \cite{chechik2020uncertainty}. Mitigating assurance weakeners involves reducing them \cite{chechik2020uncertainty}  or even completely eliminating (removing) them. In accordance with the map that Figure \ref{fig:taxonomy} depicts, we present in this section the different categories of approaches proposed to mitigate assurance weakeners. Table \ref{tab:mitigation-approaches} reports the corresponding primary studies.

\begin{table}[htbp]
\centering
\caption{Approaches for Mitigating Assurance Weakeners}
\label{tab:mitigation-approaches}
\begin{tabular}{>{\raggedright}p{3.5cm} p{10cm} }
\toprule
\textbf{References} & \textbf{Category of Approaches}  \\
\midrule
\cite{liu2012safety,sun2014rethinking,muram2018preventing,matsuno2019tackling,chechik2020uncertainty,muram2023attest,cobos2021cybersecurity,millet2023assurance,diemert2023incremental,goodenough2013eliminative,grigorova2014argument,diemert2020eliminative} & Argumentation \\
\addlinespace
\cite{nemouchi2019isabelle,rushby2013logic, rushby2014mechanized,denney2013evidence,denney2015formal,carlan2016integrated,carlan2016using,chechik2019software,viger2021lean,murphy2021validating} & Formalization \\
\addlinespace
\cite{gansch2020system} & Bayesian analysis \\
\addlinespace
\cite{schleiss2022towards,belle2023position,mcdermid2019towards} & Runtime monitoring \\
\bottomrule
\end{tabular}
\end{table}

\begin{itemize}
\item \textbf{Argumentation}: several approaches support argument decomposition, refinement, or review to reason away detected assurance weakeners or prevent them. For instance, Liu et al. \cite{liu2012safety} discuss a novel approach to analyzing failures of digital systems based on the concept of safety arguments. The safety argument of a failed system assists them in developing hypotheses concerning how the system might have failed, eliciting the evidence necessary to confirm or refute those hypotheses, documenting lessons, and developing recommendations to prevent the failure from recurring.
Sun et al. \cite{sun2014rethinking} suggest considering logical appeals in safety arguments such as Argument by Causation, Argument by Comparison, Argument from Two Sides, Argument from Authority or Expert and Argument by Eliminative Induction. These logic appeals help refine argument structures to mitigate potential logical fallacies. 

Muram et al. \cite{muram2018preventing} present an approach that validates the process models and prevents the occurrence of fallacy, specifically, the omission of key evidence in process-based argumentations. If fallacies are detected in the process models, the approach develops the recommendations to resolve them; afterwards, the process and/or safety engineers modify the process models based on the provided recommendations. Finally, the approach generates the safety argumentations (compliant with Structured Assurance Case Metamodel) from the modified process models by using model-driven engineering principles free from fallacies.
Matsuno et al. \cite{matsuno2019tackling} investigate uncertainty in implementations constructed by machine learning regarding continuous argument engineering with a granular performance evaluation over the expected operational domain. They employ an attribute testing method for evaluating an implemented model in terms of explicit specification. 
Chechik et al. \cite{chechik2020uncertainty} suggest addressing uncertainty by modifying the design, the system operation, and the safety argument and producing additional evidence to increase confidence. They also propose to support argument weakening and evidence composition (i.e., provide pieces of evidence of higher quality by composing them to obtain evidence that yields lower uncertainty levels). 

A particular form of argumentation is called \textbf{\textit{Eliminative Argumentation}} \cite{goodenough2015eliminative}. Eliminative argumentation involves reasoning away argument uncertainties by relying on \textit{defeaters} that challenge arguments \cite{muram2023attest}. Cobos et al. \cite{cobos2021cybersecurity} present a method that combines Attack-Defence Trees used in cybersecurity risk assessments with eliminative argumentation techniques. This combination is showcased within a GSN-based framework, aiming to establish a more targeted risk assurance case for automotive cybersecurity. Millet et al. \cite{millet2023assurance} employ the eliminative argumentation approach to address defeaters. They opted for EA because it distinctly allows for the articulation (and handling) of uncertainties regarding the soundness of the argument by incorporating defeaters. Diemert et al. \cite{diemert2023incremental} introduce an incremental assurance approach. Their suggested "syntactic pattern" clearly details the manner in which distinct pieces of evidence, or assertions validated by such evidence, can be integrated to fortify trust in addressing a defeater within an EA discourse.

\textbf{\textit{ Eliminative induction}} \cite{goodenough2015eliminative} is a particular form of induction that allows building confidence in arguments when performing eliminative argumentation. Some approaches rely on the Baconian philosophy of eliminative induction and the use of defeasible reasoning to generate assurance weakeners and reason them away. Goodenough et al. \cite{goodenough2013eliminative}, Grigorova and Maibaum \cite{grigorova2014argument}, as well as Diemert et al. \cite{diemert2020eliminative} suggest using eliminative induction and defeasible reasoning for mitigating defeaters (doubts). In eliminative induction, a claim is justified only to the extent that reasons for doubting its truth have been eliminated. As grounds for doubt are eliminated (through evidence or argument), confidence in the truth of the claim increases. In addition, they use defeasible reasoning concepts to generate reasons for doubting the truth of a claim (Potential reasons for doubt are called defeaters) and reason these doubts away. These reasons are usually documented within a confidence map that allows the explicit reasoning about sources of doubt in an argument \cite{chechik2019software} and eliminating these doubts.

\item \textbf{Formalization}: Other approaches rely on formal methods to enable verification of argument consistency and well-formedness. This is the case of Nemouchi et al. \cite{nemouchi2019isabelle}, who suggest using formalization in a machine-checked logic to enable verification of consistency and Well-formedness of assurance cases which might suffer from logical fallacies and inadequate evidence. Rushby \cite{rushby2013logic, rushby2014mechanized} suggests using formal verification systems, which provide tools that can be adapted to represent, analyze, and explore the logic of the case, thereby mitigating logic doubt. Moreover, to mitigate epistemic doubt, they suggest building models in logic that describe the elements of our knowledge (e.g., the behaviour of the environment), and these models should be described as systems of constraints. These can be explored and validated using tools based on SMT (satisfiability modulo theories) solvers.  

Denney and Pai \cite{denney2013evidence} discuss that an evidence argument including additional information, such as the tool-specific claims made and the assumptions/reasoning underlying the formal method/tool, provides a richer view of tool-based evidence. This argument can be independently scrutinized to mitigate the assurance deficit and improve the assurance that can be provided. The evidence argument could be created from the output of a theorem prover-based verification tool and then integrated automatically into the wider assurance case. Denney et al.  \cite{denney2015formal} proposed hierarchical safety cases, \textit{hicases}, to aid the comprehension of safety case argument structures. The broad goal is to make safety cases amenable to formal analysis, thereby providing greater assurance and mitigating any potential deficit. 

Cârlan et al. \cite{carlan2016integrated} suggest using integrated formal methods in verification activities since deficits in the verification process may negatively impact the confidence of verification results. They use integrated formal methods as evidence in assurance cases, which are used to certify safety-critical systems. They first present two workflows that employ integrated formal methods – code review workflow and code coverage workflow – corresponding to two of the most important activities of the verification phase. 
Cârlan et al.  \cite{carlan2016using} also proposed an assurance case pattern which addresses the disciplined use of successful but possibly incomplete verification results obtained through C-level bounded model checking as evidence in certification. They propose a strategy to express confidence in incomplete verification results by complementing them with classical testing and mitigating the assurance deficits with additional tests.

Chechik et al. \cite{chechik2019software} propose that integrating a certain level of formality within assurance cases can significantly enhance its validity, highlight hidden uncertainties, prevent errors, and also bolster its modularity, adaptability, and reusability. Additionally, they recommend addressing specific assurance deficits through a combination of verification methods. Viger et al. \cite{viger2021lean} and Murphy et al. \cite{murphy2021validating}  suggest using the interactive theorem prover Lean to bridge the gap between safety arguments and rigorous model-based reasoning. They generate formal, model-based, machine-checked AC arguments, taking advantage of the traceability between model and safety artifacts and mitigating errors that could arise from manual argument assessment.

\item \textbf{Bayesian analysis}: Some approaches employ a Bayesian network approach coupled with evidence theory to support assurance weakeners mitigation. Gansch et al. \cite{gansch2020system} recommend adopting fault tree analysis (FTA) to address uncertainties. FTA is a graphical representation rooted in Boolean fault propagation, assisting in pinpointing system vulnerabilities such as single-point flaws.
Although FTA is widely recognized and frequently employed, it's not without limitations. Especially for autonomous systems, FTA's emphasis on failures restricts its ability to consider human aspects of the system's standard performance.
To address FTA's limitations concerning diverse uncertainties, the authors propose a new approach that utilizes evidence theory coupled with Bayesian networks (BN). Evidence theory encompasses various uncertainties, including epistemic and ontological.

\item \textbf{Runtime monitoring}: Other approaches dynamically monitor assurance weakeners at runtime to provide continuous assurance. To provide continuous (dynamic) assurance, Schleiss et al. \cite{schleiss2022towards} advocate for adopting real-time monitoring techniques. These methods allow a system to detect its vulnerability to unforeseen hazardous conditions autonomously. Once identified, the system can swiftly initiate safety measures and embark on a redesign and updating phase to guarantee its continued safety. Belle et al. \cite{belle2023position} propose using DACs (Dynamic Assurance Cases) to support real-time safety assurance in autonomous driving systems. They advocate for addressing aleatory uncertainty during the design phase and relying on Machine Learning to manage epistemic uncertainty at runtime (during system operation). McDermid et al.  \cite{mcdermid2019towards} propose a framework for safety assurance that uses machine learning to provide evidence for a system safety case and thus enables the safety case to be updated dynamically as system behaviour evolves and mitigates assurance deficits. 
\end{itemize}

 \fbox{%
  \parbox{\dimexpr\linewidth-2\fboxsep-2\fboxrule\relax}{
    \textbf{Key Takeaway:}\\
Findings from our analysis suggest that argumentation methodologies, particularly Eliminative Argumentation (EA), have emerged as the foremost and most frequently employed techniques for mitigating assurance weakeners. These techniques offer a systematic method of addressing and eliminating weaker arguments, thereby strengthening the overall assurance stance. Along with this, runtime monitoring mechanisms—exemplified by the implementation of Dynamic Assurance Cases and the integration of machine learning to furnish real-time evidence—appear to be a potent strategy for mitigating uncertainties. These innovative approaches use  machine learning not only support the effectiveness of runtime assurance but also open up new avenues for research. Given these findings, it is imperative for the research community to dedicate more efforts on utilizing machine learning methodologies in the realm of assurance weakener mitigation.

}
}

\section{Discussion}
\subsection{Limitations}
We adopt the framework that Wohlin et al.  \cite{wohlin2012experimentation},  and Zhou et al. \cite{zhou2016map} propose to discuss the limitations of our systematic review.
 
\subsubsection{Internal validity}
In our selection and data extraction processes, the potentially subjective analysis of the reviewers could introduce bias and the risk of missing relevant results. To minimize these potential issues, a rigorous approach was adopted. Two reviewers were involved in the selection of primary studies and data extraction. During selection, one researcher initially chose the papers, while the other conducted random sampling to ensure relevance with the eligibility criteria. Data extraction was also conducted by both reviewers, with the first reviewer covering all papers and the second reviewer extracting data from half of them.
Reviewers discussed any disagreements during meetings to reach a consensus.


\subsubsection{Construct validity}
While our database-driven search employed a query that may not have been initially exhaustive, we took measures to enhance its completeness. Through iterative refinement based on the existing literature, we progressively developed a query that effectively captured a substantial body of relevant work. Moreover, the limitation of searching within five digital libraries may have introduced the risk of overlooking relevant studies. To address this limitation, we employed a snowballing technique to identify additional studies, especially those not present in the databases of the five libraries. This ensures the completeness of our search process.

\subsubsection{Conclusion validity}
By adhering to systematic review guidelines and transparently reporting our methodology, we have ensured the reproducibility of our study. Researchers with access to the same libraries and resources we mention in this study can replicate our study and expect to obtain similar results. This transparency enhances the overall reliability and validity of our research. It's worth noting that they may uncover additional findings, especially for papers published after the completion of our search. For instance, when it comes to studies published in 2023, we searched for studies published between January 2023 and the beginning of October 2023. So, our search may miss the few studies published between October and December of the same year, if there are any. Thus, surveying studies published that year may have yielded a partial and potentially biased overview of the scientific contributions made in 2023 regarding the surveyed topic.

\subsection{Implications Of The Results For Practice, Policy And Future Research}
The outcomes of our systematic mapping study allow us to conclude that several future directions
still need to be explored to better deal with assurance weakeners and increase the assurance of systems. We discuss these directions in the remainder of this section.

\subsubsection{Improvement of existing Assurance Case Notations to better represent Assurance Weakeners}
The findings of our systematic review suggest that we may need to extend existing graphical notations (e.g., GSN, CAE) to better present assurance weakeners. 

In Section \ref{repWeakenerS}, we specifically demonstrated that only \textbf{argument uncertainty} is depicted by the current notations, while aleatory, epistemic, and ontological uncertainties lack explicit representation. Such omissions complicate modeling-level reasoning and the formulation of effective solutions for their detection and mitigation. One might question the logic of such representations.

Besides, as Chechik et al.  \cite{chechik2019software} point out, arguments may lack rigour. Such arguments usually miss several critical properties, including completeness, independence, relevance, or a clear statement of assumptions. So, it may be interesting to analyze the specifications of existing assurance case notations to determine if they need to be extended to better support these properties and provide a better representation of assurance weakeners in arguments, when applicable. Besides, explicit modelling and management of uncertainty in evidence, specifications and assumptions, and the clear justification of each step of the development of a system, can go a long way toward making arguments valid, reusable, and generally helpful in helping produce high-quality software systems.

Moreover, extending well-established notations could help support a better representation of 1) different types of weakeners; 2) relationships between assurance weakeners and existing notations concepts, allowing to representation of the traditional argumentation structure made of claims, arguments and evidence; 3) assurance weakeners decorators. Such extensions could facilitate the communication of concerns and limitations about the system being audited \cite{diemert2020eliminative}.

Still, in Section \ref{repWeakenerS}, we showed that SACM may be superior to other notations, especially when it comes to representing weakeners. This is because it is a standard that unifies GSN and CAE \cite{nemouchi2019isabelle} and supports all types of defeaters and their relationships. Consequently, there should be a heightened emphasis on further research on SACM that will probably witness a greater adoption in the upcoming years.

\subsubsection{Automatic Generation of Assurance Weakeners }
Our findings from Section \ref{detWeakenerS} show that there is a lack of a holistic approach for identifying every type of assurance weakeners. Hence, one of the promising directions for future research in the field of assurance cases may be the development of tools for the automatic generation (elicitation) and/or detection of assurance weakeners. These tools can significantly enhance the verification of assurance cases by ensuring their structure is correct and properly addresses potential assurance weakeners that may have been overlooked when constructing assurance cases. Leveraging Artificial Intelligence techniques, and more specifically cutting-edge technologies such as Generative Pre-trained Transformers (e.g., GPT-4), holds great potential for automatically generating assurance weakeners. The large language models they support can be trained to understand and generate various types of defeaters. This approach aligns with recent advancements in natural language processing and generation, as demonstrated in the work by Chen et al.  \cite{chen2023use}. The latter relies on GPT-4 to create goal models. Additionally, recent preliminary work by Viger et al. \cite{viger2023supporting} elaborate their vision on the possibility to use Generative AI to identify defeaters in assurance cases. 
Such approaches may have the potential to significantly improve the accuracy of defeaters identification. The capacity of GAI to simulate a myriad of scenarios and reason creatively gives it an edge in identifying potential vulnerabilities or gaps, making it an invaluable tool in improving the reliability and robustness of assurance cases.

\subsubsection{Automatic Mitigation of Assurance Weakeners}
Our findings from Section \ref{mitWeakenerS} show that a  promising approach for automatically mitigating assurance weakeners is using the Eliminative Argumentation (EA) algorithm \cite{goodenough2015eliminative,cobos2021cybersecurity,millet2023assurance}. EA offers a conceptual structure for formulating an argument and evaluating trust in that argument using the idea of defeasible reasoning. In this method, claims are continually questioned, and as doubts regarding a claim are removed, confidence in that claim strengthens \cite{diemert2020eliminative}. Automating such an algorithm into the assurance case development process can proactively mitigate potential weakeners. Furthermore, as discussed in Section \ref{mitWeakenerS} Key Takeaway, it may be interesting to further explore the use of machine learning techniques to mitigate assurance weakeners. In this regard, when trained on a sufficiently large dataset of assurance cases, machine learning models can suggest potential patterns to mitigate each detected assurance weakeners based on patterns from past data. Coupled with NLP, these models can also understand the context and semantics of the assurance case narratives, potentially leading to a more accurate detection of assurance weakeners.

\section{Other Information}
\subsection{Support}
The start-up grant of the second author funded this research.

\subsection{Competing interests}
No member of the research team has disclosed any potential conflicts of interest, whether real or perceived, in relation to the subject of this study.

\section{Conclusion}
In the sphere of cyber-physical systems (CPSs), assurance arguments stand out as paramount. These CPSs, encompassing healthcare to aviation, including critical systems like autonomous vehicles, hold immense responsibility, especially where human safety is involved. The complexity of these systems, particularly when integrating machine learning components, underscores the challenge of ensuring they function dependably even under unpredictable circumstances. Assurance weakeners pose a significant risk here, potentially eroding our confidence in these systems.
This research, structured around the PRISMA guideline, offers a profound understanding of these weakeners. By categorizing them—terms like 'uncertainty' and 'doubt' come into play—we provide a clearer picture of their intricacies and potential implications. Beyond identification, our focus extends to exploring strategies for their mitigation, all aimed at strengthening the foundation of assurance cases.
Looking to the future, we're envisioning a cohesive framework for assurance cases. This framework, informed by our categorization, is set to promote automated processes to represent, identify and mitigate weakeners more effectively. Additionally, our intent is to integrate this with the DevCase tool \cite{Wang2023DevCase}, enhancing its practical application.
An essential highlight from our findings is that SACM may be the best specification to represent assurance weakeners. It is a potential specification for framing structured arguments, demonstrating a distinct edge over other notations due to its adaptability and depth. Conclusively, this research holds profound implications for the world of CPSs. As these systems increasingly make their case to regulatory bodies using assurance cases, our findings emphasize the necessity for fresh strategies, ensuring these vital systems' unwavering reliability.

\appendix

\section{Queries employed to search Each Database}
Table \ref{tab:queries} report the query(ies) we searched in each of the five databases.
We used the Publish or Perish tool to search for the keywords for the query on Google Scholar. Google Scholar has a 256-character limit for searches \cite{google_scholar_limit}, and since our query had more characters than the limit, we had to split it into multiple smaller queries. 
 
\label{queries}
\begin{longtable}[htbp]{p{2cm}p{8cm}p{4.5cm}}
\caption{Queries searched in each database}
\label{tab:queries}\\
\toprule
\textbf{Database name} & \textbf{Query(ies)} & \textbf{Database parameters used for the search} \\
\midrule
\endfirsthead

\multicolumn{3}{c}%
{{\tablename\ \thetable{} -- continued from previous page}} \\
\toprule
\textbf{Database name} & \textbf{Query(ies)} & \textbf{Database parameters used for the search} \\
\midrule
\endhead

\bottomrule \multicolumn{3}{r}{{Continued on next page}} \\ \bottomrule
\endfoot

\bottomrule
\endlastfoot

Google Scholar & \begin{itemize}
  \item \textbf{Query 1:} ("assurance deficits" OR “false assurance” OR "defeaters") AND (“assurance case” OR “safety case” OR “trust case” OR “dependability case” OR “reliability case” OR “security case” OR “availability case”)

  \item \textbf{Query 2:} ("counter evidence" OR "counter-argument" OR "fallacies") AND (“assurance case” OR “safety case” OR “trust case” OR “dependability case” OR “reliability case” OR “security case” OR “availability case”)

  \item \textbf{Query 3:} ("aleatory uncertainty" OR "aleatoric uncertainty" OR "epistemic uncertainty") AND (“assurance case” OR “safety case” OR “trust case” OR “dependability case” OR “reliability case” OR “security case” OR “availability case”)

  \item \textbf{Query 4:} ("uncertainty reasoning" OR “uncertainty elicitation” OR “informal semantics” OR “doubt”) AND (“assurance case” OR “safety case” OR “trust case” OR “dependability case” OR “reliability case” OR “security case” OR “availability case”)
\end{itemize}

& Keyword, 2012-2023 \\

Scopus &
TITLE-ABS-KEY(("assurance deficits" OR “false assurance” OR "defeaters" OR "counter evidence" OR "counter-argument" OR "fallacies" OR "aleatory uncertainty" OR "aleatoric uncertainty" OR "epistemic uncertainty" OR "uncertainty reasoning" OR “uncertainty elicitation” OR “informal semantics” OR “doubt”) AND (“assurance case” OR “safety case” OR “trust case” OR “dependability case” OR “reliability case” OR “security case” OR “availability case”)) AND PUBYEAR AFT 2011
 &  Advanced search on title/abstract/keyword, Published after 2011\\
 \hline

IEEE Xplore & 
("assurance deficits" OR “false assurance” OR "defeaters" OR "counter evidence" OR "counter-argument" OR "fallacies" OR "aleatory uncertainty" OR "aleatoric uncertainty" OR "epistemic uncertainty" OR "uncertainty reasoning" OR “uncertainty elicitation” OR “informal semantics” OR “doubt”) 
\newline
AND 
\newline
(“assurance case” OR “safety case” OR “trust case” OR “dependability case” OR “reliability case” OR “security case” OR “availability case”)
& Command search on metadata\\
ACM & 
("assurance deficits" OR “false assurance” OR "defeaters" OR "counter evidence" OR "counter-argument" OR "fallacies" OR "aleatory uncertainty" OR "aleatoric uncertainty" OR "epistemic uncertainty" OR "uncertainty reasoning" OR “uncertainty elicitation” OR “informal semantics” OR “doubt”) 
\newline
AND 
\newline
("assurance case" OR “safety case” OR “trust case” OR “dependability case” OR “reliability case” OR “security case” OR “availability case”)
& Advanced search on the
title and abstract \\
\hline

Engineering Village & 
("assurance deficits" OR “false assurance” OR "defeaters" OR "counter evidence" OR "counter-argument" OR "fallacies" OR "aleatory uncertainty" OR "aleatoric uncertainty" OR "epistemic uncertainty" OR "uncertainty reasoning" OR “uncertainty elicitation” OR “informal semantics” OR “doubt”) 
\newline
AND 
\newline
(“assurance case” OR “safety case” OR “trust case” OR “dependability case” OR “reliability case” OR “security case” OR “availability case”)
& Search on the
subject, title and abstract \\

\end{longtable}

\section{Study Characteristics}
\label{studychar}
{\footnotesize
\begin{longtable}[htbp]{p{0.7cm}|p{3cm}|p{6cm}|p{2.2cm}|p{2.7cm}}
\caption{List of Studies}\\
\label{tab:studies}\\
\hline
 \textbf{No.} &
 \textbf{Authors (Publication Year)} & \textbf{Study Title} & \textbf{Venue} & \textbf{Type of Search} \\
\hline
\endfirsthead

\multicolumn{5}{c}%
{{\bfseries \tablename\ \thetable{} -- continued from previous page}} \\
\hline
\textbf{No.} &
\textbf{Authors (Publication Year)} & \textbf{Study Title} & \textbf{Venue} & \textbf{Type of Search} \\
\hline
\endhead

\hline \multicolumn{5}{|r|}{{Continued on next page}} \\ \hline
\endfoot

\hline \hline
\endlastfoot

1 & Liu et al. (2012) \cite{liu2012safety}  & A safety-argument based method to predict system failure & ICPHM & Database-driven \\ \hline

2 & Denney and Pai. (2013) \cite{denney2013evidence}  & Evidence arguments for using formal methods in software certification & ISSRE & Database-driven \\ \hline

3 & Goodenough et al. (2013) \cite{goodenough2013eliminative}  & Eliminative induction: A basis for arguing system confidence & ICSE & Database-driven \\
\hline

4 & Rushby (2013) \cite{rushby2013logic}  & Logic and epistemology in safety cases & SAFECOMP & Database-driven \\
\hline

5 & Yamamoto et al. (2013) \cite{yamamoto2013evaluation}& An evaluation of argument patterns to reduce pitfalls of applying assurance case & ASSURE & Database-driven  \\
\hline

6 & Graydon (2014) \cite{graydon2014towards} & Towards a clearer understanding of context and its role in assurance argument confidence & SAFECOMP & Database-driven \\
\hline

7 & Grigorova and Maibaum (2014) \cite{grigorova2014argument}  & Argument evaluation in the context of assurance case confidence modeling & ISSRE & Database-driven  \\
\hline

8 & Rushby (2014) \cite{rushby2014mechanized} & Mechanized support for assurance case argumentation & JSAI & Database-driven  \\
\hline

9 & Sun et al. (2014) \cite{sun2014rethinking} & Rethinking of Strategy for Safety Argument Development & SAFECOMP & Database-driven \\
\hline

10 & Takai and Kido (2014) \cite{takai2014supplemental} & A supplemental notation of GSN aiming for dealing with changes of assurance cases & ISSRE & Database-driven \\
\hline

11 & Bandur and McDermid (2015) \cite{bandur2015informing} & Informing assurance case review through a formal interpretation of GSN core logic & SAFECOMP & Database-driven \\
\hline

12 & Denney et al. (2015) \cite{denney2015dynamic} & Dynamic safety cases for through-life safety assurance & ICSE & Database-driven \\
\hline

13 & Denney et al. (2015) \cite{denney2015formal} & Formal foundations for hierarchical safety cases & HASE & Database-driven \\
\hline

14 & Groza et al. (2015) \cite{groza2015formal} & A formal approach for identifying assurance deficits in unmanned aerial vehicle software & ICSEng & Database-driven \\
\hline

15 & Jarz\k{e}bowicz and Wardzi\'{n}ski (2015)  \cite{jarzbowicz2015integrating}  & Integrating confidence and assurance arguments & IET &  Database-driven \\
\hline

16 & Cârlan et al. (2016) \cite{carlan2016integrated}  & Integrated formal methods for constructing assurance cases & ISSRE & Database-driven \\
\hline

17 & Cârlan et al. (2016) \cite{carlan2016using} & On using results of code-level bounded model checking in assurance cases & SAFECOMP & Database-driven \\
\hline

18 & Yuan et al. (2016) \cite{yuan2016automatically} & Automatically detecting fallacies in system safety arguments & PRIMA & Database-driven \\
\hline

19 & Muram et al. (2018) 
\cite{muram2018preventing} & Preventing omission of key evidence fallacy in process-based argumentations & QUATIC & Database-driven \\
\hline

20 & Chechik et al. (2019) \cite{chechik2019software}  & Software Assurance in an Uncertain World & FASE & Snowballing \\
\hline

21 & Matsuno et al. (2019) \cite{matsuno2019tackling} & Tackling Uncertainty in Safety Assurance for Machine Learning: Continuous Argument Engineering with Attributed Tests & SAFECOMP & Snowballing \\
\hline

22 & McDermid et al. (2019) \cite{mcdermid2019towards} & Towards a Framework for Safety Assurance of Autonomous Systems & AISafety & Snowballing  \\
\hline

23 & Nemouchi et al. (2019) \cite{nemouchi2019isabelle} & Isabelle/SACM: Computer-assisted assurance cases with integrated formal methods & iFM & Database-driven \\
\hline

24 & Chechik et al. (2020) \cite{chechik2020uncertainty} & Uncertainty, modeling and safety assurance: towards a unified framework & VSTTE & Database-driven  \\
\hline

25 & Diemert et al. (2020) \cite{diemert2020eliminative} & Eliminative Argumentation for Arguing System Safety-A Practitioner's Experience & SysCon & Database-driven   \\
\hline

26 & Gansch et al. (2020) \cite{gansch2020system} & System theoretic view on uncertainties & DATE & Database-driven  \\
\hline

27 & Jarzębowicz and Markiewicz (2020) \cite{jarzkebowicz2020representing} & Representing process characteristics to increase confidence in assurance case arguments & DepCoS & Database-driven \\

\hline

28 & Selviandro et al. (2020) \cite{selviandro2020visual} & A visual notation for the representation of assurance cases using sacm & IMBSA & Database-driven  \\

\hline

29 & Foster et al. (2021) \cite{foster2021integration} & Integration of Formal Proof into Unified Assurance Cases with Isabelle/SACM & FAC & Database-driven \\
\hline

30 & Murphy et al. (2021) \cite{murphy2021validating} & Validating safety arguments with lean & SEFM & Database-driven \\
\hline

31 & Viger et al. (2021) \cite{viger2021lean} & A lean approach to building valid model-based safety arguments & MODELS & Database-driven \\
\hline

32 & Cobos et al. (2021) \cite{cobos2021cybersecurity} & Cybersecurity Assurance Challenges for Future Connected and Automated Vehicles & ESREL & Snowballing \\
\hline

33 & Schleiss et al. (2022) \cite{schleiss2022towards}  & Towards Continuous Safety Assurance for Autonomous Systems & ICSRS & Database-driven \\
\hline

34 & Belle et al. (2023) \cite{belle2023position} & Position paper: a vision for the dynamic safety assurance of ML-enabled autonomous driving systems & REW & Database-driven \\
\hline

35 & Diemert et al. (2023) \cite{diemert2023incremental} & Incremental Assurance Through Eliminative Argumentation & JSS & Database-driven \\
\hline

36 & Menghi et al. (2023) \cite{menghi2023assurance} & Assurance case development as data: A manifesto & ICSE-NIER & Database-driven \\
\hline

37 & Millet et al. (2023) \cite{millet2023assurance} (2023) & Assurance Case Arguments in the Large: The CERN LHC Machine Protection System & SAFECOMP & Database-driven \\
\hline

38 & Muram and Javed (2023) \cite{muram2023attest} & ATTEST: Automating the review and update of assurance case arguments & JSA & Database-driven \\

\hline

39 & Murugesan et al. (2023) \cite{murugesan2023semantic} & Semantic Analysis of Assurance Cases using s(CASP) & ICLP & Database-driven 

\end{longtable}
}

\section{Venue Names and Acronyms}

\label{acronyms}
{\footnotesize
\begin{longtable}[htbp]{p{0.6\textwidth}|p{0.3\textwidth}}
\caption{Venue Names and Acronyms} \\
\hline
\textbf{Venue Name} & \textbf{Acronym} \\
\hline
\endfirsthead
\multicolumn{2}{c}%
{\tablename\ \thetable\ -- \textit{Continued from previous page}} \\
\hline
\textbf{Venue Name} & \textbf{Acronym} \\
\hline
\endhead
\hline \multicolumn{2}{r}{\textit{Continued on next page}} \\
\endfoot
\hline
\endlastfoot
IEEE International Symposium on Software Reliability Engineering Workshops & ISSRE \\
\hline
Prognostics and System Health Management Conference & ICPHM \\
\hline
Fundamental Approaches to Software Engineering & FASE \\
\hline
International Conference on Software Engineering and Formal Methods & SEFM \\
\hline
Formal Aspects of Computing & FAC \\
\hline
ACM/IEEE International Conference on Model Driven Engineering Languages and Systems & MODELS \\
\hline
Verified Software. Theories Tools and Experiments & VSTTE \\
\hline
Design Automation \& Test in Europe Conference \& Exhibition & DATE \\
\hline
International Conference on Dependability and Complex Systems & DepCoS \\
\hline
IEEE International Systems Conference & SysCon \\
\hline
International Symposium on Model-Based Safety and Assessment & IMBSA \\
\hline
Artificial Intelligence Safety & AISafety \\
\hline
Integrated Formal Methods & iFM \\
\hline
International Workshop on Assurance Cases for Software-Intensive Systems & ASSURE \\
\hline
International Conference on the Quality of Information and Communications Technology & QUATIC \\
\hline
Principles and Practice of Multi-Agent Systems & PRIMA \\
\hline
IET System Safety and Cyber-Security Conference & IET \\
\hline
International Symposium on High Assurance Systems Engineering & HASE \\
\hline
International Conference on Software Engineering & ICSE \\
\hline
International Conference on Systems Engineering & ICSEng \\
\hline
JSAI International Symposium on Artificial Intelligence & JSAI \\
\hline
International Conference on System Reliability and Safety & ICSRS \\
\hline
European Safety and Reliability Conference & ESREL \\
\hline
International Requirements Engineering Conference Workshops & REW \\
\hline
Journal of System Safety & JSS \\
\hline
International Conference on Software Engineering: New Ideas and Emerging Results & ICSE-NIER \\
\hline
Journal of Systems Architecture & JSA\\
\hline
International Conference on Logic Programming & ICLP \\

\hline
\end{longtable}
}


\bibliographystyle{ieeetr}
\bibliography{references}

\end{document}